\documentclass[10pt,journal,compsoc]{IEEEtran}

\ifCLASSINFOpdf
\else
\fi

\usepackage{algorithmic}
\usepackage{array}
\usepackage{textcomp}
\usepackage{stfloats}
\usepackage{verbatim}
\usepackage{graphicx}
\usepackage{booktabs}
\usepackage{fleqn}
\usepackage{multirow}
\usepackage{amssymb}
\usepackage{amsmath,amsfonts}
\usepackage{amsthm}
\usepackage{graphicx}
\usepackage{cite}
\usepackage{array}
\usepackage{color}
\usepackage{bm}
\usepackage{subfigure}
\usepackage{makecell}
\usepackage[linesnumbered,algoruled,boxed,lined,noend]{algorithm2e}
\usepackage{tabularx}
\usepackage{enumerate}
\usepackage{enumitem}
\usepackage[normalem]{ulem}
\usepackage{url}
\usepackage{microtype}
\usepackage{balance}
\usepackage{titlesec} 
\titlespacing{\section}{0pt}{1.3ex}{1.3ex}
\titlespacing{\subsection}{0pt}{1.3ex}{1.3ex}
\titlespacing{\subsubsection}{0pt}{1.3ex}{1.3ex}

\newtheorem{Theorem}{Theorem}
\newtheorem{Definition}{Definition}

\newtheorem{Lemma}{Lemma}

\allowdisplaybreaks[4]
\begin{document}

\title{UAV Swarm-enabled Collaborative Post-disaster Communications in Low Altitude Economy via a Two-stage Optimization Approach}

\author{Xiaoya Zheng,
        Geng Sun,~\IEEEmembership{Senior Member,~IEEE,}
        Jiahui Li,~\IEEEmembership{Student Member,~IEEE,}\\
        Jiacheng Wang,
        Qingqing Wu,~\IEEEmembership{Senior Member,~IEEE,}
        Dusit Niyato,~\IEEEmembership{Fellow,~IEEE,}\\
        and Abbas Jamalipour,~\IEEEmembership{Fellow,~IEEE}
        
        \thanks{This study is supported in part by the National Natural Science Foundation of China (62272194, 62172186), in part by the Science and Technology Development Plan Project of Jilin Province (20230201087GX). (Corresponding authors: Geng Sun and Jiahui Li).   
        \par Xiaoya Zheng and Jiahui Li are with the College of Computer Science and Technology, Jilin University, Changchun 130012, China (E-mails: \protect\url{xiaoya257248@foxmail.com}, lijiahui0803@foxmail.com).
        \par Geng Sun is with the College of Computer Science and Technology, Jilin University, Changchun 130012, China, and also with the Key Laboratory of Symbolic Computation and Knowledge Engineering of Ministry of Education, Jilin University, Changchun 130012, China. He is also with the College of Computing and Data Science, Nanyang Technological University, Singapore 639798 (E-mail: sungeng@jlu.edu.cn).
        \par Jiacheng Wang and Dusit Niyato are with the College of Computing and Data Science, Nanyang Technological University, Singapore 639798 (E-mail: \protect\url{jiacheng.wang@ntu.edu.sg, dniyato@ntu.edu.sg}).
        \par Qingqing Wu is with the Department of Electronic Engineering, Shanghai Jiao Tong University, Shanghai 200240, China (E-mail: \protect\url{qingqingwu@sjtu.edu.cn}).
        \par Abbas Jamalipour is with the School of Electrical and Computer Engineering, The University of Sydney, Sydney, NSW 2006, Australia (E-mail: \protect\url{a.jamalipour@ieee.org}).
        }
}

\IEEEtitleabstractindextext{
\begin{abstract}
The low-altitude economy (LAE), as a new economic paradigm, plays an indispensable role in cargo transportation, healthcare, infrastructure inspection, and especially post-disaster communication. Specifically, unmanned aerial vehicles (UAVs), as one of the core technologies of the LAE, can be deployed to provide communication coverage, facilitate data collection, and relay data for trapped users, thereby significantly enhancing the efficiency of post-disaster response efforts. However, conventional UAV self-organizing networks exhibit low reliability in long-range cases due to their limited onboard energy and transmit ability. Therefore, in this paper, we design an efficient and robust UAV-swarm enabled collaborative self-organizing network to facilitate post-disaster communications. Specifically, a ground device transmits data to UAV swarms, which then use collaborative beamforming (CB) technique to form virtual antenna arrays and relay the data to a remote access point (AP) efficiently. Then, we formulate a rescue-oriented post-disaster transmission rate maximization optimization problem (RPTRMOP), aimed at maximizing the transmission rate of the whole network. Given the challenges of solving the formulated RPTRMOP by using traditional algorithms, we propose a two-stage optimization approach to address it. \emph{In the first stage}, the optimal traffic routing and the theoretical upper bound on the transmission rate of the network are derived. \emph{In the second stage}, we transform the formulated RPTRMOP into a variant named V-RPTRMOP based on the obtained optimal traffic routing, aimed at rendering the actual transmission rate closely approaches its theoretical upper bound by optimizing the excitation current weight and the placement of each participating UAV via a diffusion model-enabled particle swarm optimization (DM-PSO) algorithm. Simulation results show the effectiveness of the proposed two-stage optimization approach in improving the transmission rate of the constructed network, which demonstrates the great potential for post-disaster communications. Moreover, the robustness of the constructed network is also validated via evaluating the impact of two unexpected situations on the system transmission rate.
\end{abstract}

\begin{IEEEkeywords}
Unmanned aerial vehicles (UAVs), post-disaster communications, collaborative beamforming (CB), traffic routing design, UAV swarm optimization.
\end{IEEEkeywords}
}

\maketitle
%

\IEEEdisplaynontitleabstractindextext
\IEEEpeerreviewmaketitle

\IEEEraisesectionheading{
\section{Introduction}
\label{sec: Introduction}
}

\IEEEPARstart{L}{ow-altitude} economy (LAE), centered on low-altitude airspace, serves as a burgeoning and comprehensive economic form of future economic activities, carrying out various low-altitude flight activities below 1000 meters in altitude~\cite{jiang20236g},~\cite{yang2024case}. LAE integrates advanced technologies such as unmanned aerial vehicles, satellite communication, artificial intelligence, network communication, and unmanned traffic management systems to enable the efficient integration of applications across various fields, including logistics, emergency response, and urban management~\cite{liao2024benefits}. The synergy among these technologies not only drives the development of the low-altitude economy but also positions it as a crucial link connecting ground-based economies with airspace resources, offering new opportunities for social and economic development as well as technological innovation~\cite{cheng2024networked}~\cite{openai2023chatgpt}.

\par Beyond the aforementioned applications, the potential of the LAE in disaster relief is immense. Natural disasters, such as earthquakes, floods, or typhoons, often lead to the paralysis of terrestrial infrastructure and the malfunction of communication systems, significantly impairing the efficiency of search and rescue operations~\cite{Wang2023multihop}. In such extreme conditions, LAE technologies, particularly unmanned aerial vehicle (UAV) communication networks, can provide essential emergency communication services to the trapped ground users. Specifically, due to their agility and mobility, UAVs can be swiftly deployed to various gathering points, where affected users converge around limited rescue resources, providing reliable wireless communication services within their coverage areas~\cite{Zhao2019}. Moreover, their high altitude facilitates line-of-sight (LoS) communication, which means that even though the distances among these gathering points are considerable, UAVs are still able to establish a self-organizing network to achieve connectivity both between gathering points and with remote access points (APs) or base stations~\cite{Yang2023sensorjournal}.

\par However, conventional UAV-enabled self-organizing networks face significant limitations. Typically, data is forwarded by a single UAV to the next hop or receiver, which leads to two challenges. First, data forwarding is restricted to a single path, which significantly limits the transmission rate. Second, if the UAV responsible for a particular hop fails, the transmission task is completely disrupted, potentially leading to the collapse of the entire network. Thus, how to establish efficient and robust self-organizing networks becomes a critical challenge in such scenarios.

\par Collaborative beamforming (CB) is an effective approach to addressing the aforementioned issues~\cite{garza2016design}. Specifically, several UAV elements can be deployed to form a virtual antenna array (VAA) and then transmit data towards the remote receiver in a cooperative manner~\cite{Sun2021}. By using CB, the communication capabilities between widely separated gathering points and with external networks can be significantly enhanced. The application of CB offers the following typical advantages. First, compared with the conventional multi-hop communication mechanism, where multiple UAVs may fly far to predefined locations, CB only requires the UAVs to fine-tune the excitation current weights and adjust their placements to achieve a beam pattern with a high-gain mainlobe, making it a cost-effective approach, especially for UAV networks with limited resources. Second, with $N_{\mathrm{U}}$ UAV elements, the formed VAA can achieve an $N_{\mathrm{U}}^2$ fold gain in received power, thereby significantly boosting communication performance~\cite{Jayaprakasam2017}. In conclusion, CB provides a plausible direction to establish a UAV-swarm enabled collaborative self-organizing network for post-disaster communications, enabling the connections between gathering points and between the gathering points and external APs.

\par Despite the advantages of CB, implementing a UAV-swarm enabled collaborative self-organizing network for post-disaster relief still faces several critical challenges. First, inappropriate traffic routing design can cause some nodes to become overloaded, leading to increased packet loss rates. Therefore, designing appropriate routing protocols is essential to ensure the integrity and reliability of post-disaster data transmission. Second, even with a designed traffic routing, UAV swarm optimization remains indispensable. By optimizing parameters such as UAV placement and resource allocation, resource utilization efficiency can be enhanced, which further reduces the communication delays and energy consumption. These challenges motivate us to explore new approaches that differ from existing literature. Different from our previous work, which focused on a single-hop architecture~\cite{sun2023uav}, this study presents a novel UAV swarm-enabled collaborative communication mechanism capable of achieving extended-distance communication in post-disaster scenarios. The main contributions are summarized as follows.

\begin{itemize}

\item \textbf{\emph{Virtual Antenna Array-based Collaborative UAV Networking for Post-disaster Communications:}} We consider a post-disaster communication scenario, and design a novel network structure in which multiple UAV swarms relay data from a ground device in the post-disaster area to a remote access point (AP) in a CB manner. Intuitively, the constructed network is a realistic and meaningful system since the deployment of multiple UAV swarms can cover a wider post-disaster area and enhance network reliability. Moreover, applying CB in UAV networks is expected to significantly enhance data transmission efficiency. To the best of our knowledge, the integration of CB with multiple UAV swarms for post-disaster communications has not yet been investigated.

\item \textbf{\emph{A Rescue-oriented Post-disaster Transmission Rate Maximization Problem Formulation:}} We formulate an optimization problem, termed rescue-oriented post-disaster transmission rate maximization optimization problem (RPTRMOP), to improve the transmission rate of the communication network by conducting traffic routing design and controlling the excitation current weights and placements of UAVs. Though the structure of RPTRMOP appears simple, it cannot be directly handled using existing optimization algorithms.
	
\item \textbf{\emph{Two-stage Optimization Approach Using Ford-Fulkerson and DM-PSO Algorithms:}} We propose an efficient two-stage optimization approach to deal with the formulated RPTRMOP. \emph{In the first stage}, the optimal traffic routing and the theoretical upper bound on the transmission rate of the system are derived through theoretical analysis and the Ford-Fulkerson algorithm~\cite{fordfulkersson2015flows}. \emph{In the second stage}, we transform the formulated RPTRMOP into a variant named V-RPTRMOP based on the obtained optimal traffic routing, and propose a diffusion model-enabled particle swarm optimization (DM-PSO) algorithm to address it by optimizing the excitation current weight and the placement of each participating UAV, so that the actual transmission rate closely approaches its theoretical upper bound.
		
\item \textbf{\emph{Performance Validation of the Collaborative UAV Networking and Two-stage Optimization Approach:}} Simulation results demonstrate the effectiveness of the proposed two-stage optimization approach in improving the transmission rate of the UAV-swarm enabled collaborative self-organizing network. Moreover, the robustness of the proposed network as well as the optimization approach is also validated through evaluating the impact of two unexpected situations on system performance.

\end{itemize}

\par The remainder of this work is organized as follows. Section \ref{sec: Related work} provides a comprehensive review of related works. Section \ref{sec: System Model and Problem Formulation} introduces the system model and formulates the optimization problem. Section \ref{sec: Stage One} and \ref{sec: Step Two} present the first and second stages in addressing the formulated optimization problem, respectively. Simulation results are presented in Section \ref{sec: Simulation Results}. Finally, Section \ref{sec: Conclusion} concludes the paper.

%
%
\section{Related Work}
\label{sec: Related work}

\noindent This section reviews works on UAV-assisted post-disaster communications, routing design and UAV network optimization, and solution mechanisms in UAV communications.

\subsection{UAV-assisted Post-disaster Communications}
\label{subsec: UAV-assisted Post-disaster Communications}

\noindent The application of UAVs for disaster response and relief has been extensively studied in the literature. For example, Tran \emph{et al.}~\cite{Tran2022UAVdatacollection} deployed a UAV-mounted BS as a relay to collect data from the latency-sensitive Internet of Things (IoT) devices in post-disaster areas and then transfer it to a ground gateway. Specifically, they aimed to maximize the number of the served IoT devices, while considering the storage capacity of UAVs and the requirements of each device. Liu \emph{et al.}~\cite{liu2018transceiver} dispatched a multi-antenna UAV to provide emergency coverage for ground devices in disaster areas, with each antenna serving a device independently. To extend UAV coverage, the ground devices that are outside of its range can connect with those within it by using the proposed shortest-path routing algorithm. However, the works above rely on only a single UAV to provide communication services in post-disaster areas, which exposes the system to significant vulnerability since a failure or damage to the UAV could paralyze the entire communication system and result in data loss. Moreover, communication overload may occur as a single UAV must handle all requests from ground devices.

\par To overcome the aforementioned limitations and further enhance the coverage and reliability of post-disaster communication systems, numerous studies have explored the application of multiple UAVs working in collaboration. For example, Shah \emph{et al.}~\cite{Shah2023UAVMEC} constructed a UAV-enabled mobile edge computing (MEC) network where UAVs are responsible for executing computational tasks offloaded by user equipments. Specifically, they aimed to optimize the system utility by optimizing user associations and other factors. Do-Duy \emph{et al.}~\cite{Do-Duy2021UAVrelay} deployed multiple UAVs as airborne relay stations to facilitate communication between a massive multi-input multi-output BS and several user clusters, focusing on the joint optimization of real-time deployment and resource allocation of the dispatched UAVs. Yao \emph{et al.}~\cite{Yao2021UAVIRS} explored the application of intelligent reflective surface-assisted UAV communication networks. Then, they introduced a novel fading channel model for the network and optimized power allocation based on this model. Moreover, Wang \emph{et al.}~\cite{Wang2022UAVVFC} studied a fog computing-based UAV system where UAVs offload intensive computational missions to vehicles with high processing capabilities, and proposed a game theory-based allocation strategy. However, due to limited transmit power, UAVs typically need to stay close to the served users to ensure high communication quality. In this case, the UAVs are required to keep flying or hovering for extended periods, leading to significant energy consumption, which is challenging for resource-constrained UAVs in disaster relief tasks. 

\begin{table*}[ht]
\centering
\caption{Summary of notations}
\label{table: notations}
\begin{tabular}{|m{2cm}|m{6cm}||m{2cm}|m{6cm}|}
\hline
\textbf{Notation} & \textbf{Description} & \textbf{Notation} & \textbf{Description} \\ \hline
$\alpha$       & The path loss component      &  $\mathbf{A}$ & The generalized adjacency matrix \\ \hline
$A$      & The rotor disc area  &  $\alpha_{1}$, $\alpha_{2}$, $\alpha_{3}$, $\alpha_{4}$   & The weighting factors to balance the four proposals of the transformed problem \\  \hline
$\alpha_{\mathrm{s}}$   & The coefficient that controls the weight of the original data and noise    & $B$   & The channel bandwidth  \\  \hline
$c$      & The speed of light  & $C_{i,j}$ &   The maximum capacity of link $<i,j> \in \mathcal{E}$ \\  \hline
$c_{i,j}$ &   The residual capacity of link $<i,j> \in \mathcal{E}$  &  $c_{\mathrm{r}}(p_{\mathrm{a}})$  &  The residual capacity of the augmenting path  \\ \hline
$c_{1}$, $c_{2}$  & The learning coefficients  &  $cd_{p}$   & The crowding distance of the $p$th agent  \\  \hline
$\mathbf{cd}$, $\mathbf{cd}^{\mathrm{sorted}}$ &  The crowding distance vector of the population and its sorted version, respectively  &  $D_{u_1,u_2}$  & The distance between two UAVs $u_1$ and $u_2$ in the same swarm  \\  \hline
$D_{\mathrm{min}}$      & The threshold of the distance between two UAVs in the same swarm  & $d_{i, j}$/$H_{i,j}$  & The total/vertical distance between nodes $i$ and $j$  \\ \hline
$E_{u}^{s}$   &  The flight energy consumption of the $u$th UAV in the $s$th swarm  & $F_s$       & The array factor of the VAA formed by the $s$th UAV swarm \\ \hline
$f_\mathrm{c}$           & The carrier frequency  &   $g$  & The gravitational acceleration \\  \hline
$G_{s}$      & The antenna gain of the VAA of the $s$th swarm  &  $G_{i}$ &  The antenna gain of node $i$  \\  \hline
$G = \{\mathcal{V}, \mathcal{E}, \mathcal{W}\}$  &  The directed weighted graph that describes the communication network  &  $G_{\mathrm{r}}$  &  The residual network of the original network $G$ \\  \hline
$G'$  & The optimized network   &  $h_{i, j}$     & The channel power gain between nodes $i$ and $j$  \\   \hline
$I_u^{s}$     & The excitation current weight of the $u$th UAV in the $s$th swarm  &  $i_{\mathrm{u}}$         & The imaginary unit  \\ \hline
$\mathbf{I}$, $\mathbf{P}$ & Decision variables of the RPTRMOP. $\mathbf{I}$, the excitation current weights, and $\mathbf{P}$, the 3D positions & $\mathbf{m}_p$, $\mathbf{m}_p^{\mathrm{sorted}}$ & The distance vector of the $p$th agent and its sorted version, respectively \\ \hline
$m_{u}$        & The mass of the $u$th UAV in the $s$th swarm  & $m$, $n$       & The channel parameters of probability LoS model  \\  \hline
$N_{\mathrm{S}}$      & The total number of UAV swarms   & $N_{\mathrm{U}}$        & The number of UAVs in each swarm \\ \hline
$N_{\mathrm{L}}$/$N_{\mathrm{L}}'$  & The number of communication links in network $G$/$G'$  &  $N_{\mathrm{pop}}$  & The population size  \\  \hline
$N_{\mathrm{N}}$  & The number of neighbors of agents  & $N_{\mathrm{P}}$ & The number of perturbed agents \\  \hline
$p$       & The phase constant   &  $P_{i}$   & The transmit power of node $i$ \\  \hline
$p_{\mathrm{t}}$        & The transmit power of a UAV or the device   & $PL_{i,j}$   & The path loss between nodes $i$ and $j$  \\  \hline
$P_{\mathrm{0}}$        & The blade profile power  & $P_{\mathrm{I}}$     & The induced power in hovering status  \\  \hline
$\mathbb{P}_{i,j}^{\mathrm{LoS}}$/ $\mathbb{P}_{i,j}^{\mathrm{NLoS}}$  & The LoS/NLoS probability between nodes $i$ and $j$  &  $p_{\mathrm{a}}$  & The augmenting path in the residual network $G_{\mathrm{r}}$  \\  \hline
$p_{\mathrm{max}}$/$p_{\mathrm{min}}$  & The maximum/minimum proportion of the perturbed agents  &   $\textbf{q}_{\mathrm{d}}$/$\textbf{q}_{\mathrm{a}}$/$\textbf{q}_{u}^{s}$    &  The locations of the ground device/the AP/the $u$th UAV in the $s$th swarm  \\  \hline
$\mathbb{R}_s^3$     & The movement area of UAVs in the $s$th swarm &  $R_{i, j}$     & The transmission rate between nodes $i$ and $j$   \\  \hline
$s$, $\rho$   & The air density and rotor solidity, respectively  & $\mathcal{S}_{s}$    & The $s$th UAV swarm    \\  \hline
$T_{\mathrm{max}}$ & The number of maximum iterations &  $\mathcal{U}$       & The set of all UAV swarms  \\  \hline
$v_{\mathrm{b}}$        & The tip speed of the rotor blade   &  
$v_{\mathrm{m}}$   & The mean rotor induced velocity in hovering  \\ \hline
$v_{u}$       & The velocity of the $u$th UAV in the $s$th swarm  &
$w(\theta,\phi)$     & The magnitude of the far-field beam pattern of a UAV  \\  \hline
$\omega$  & The inertia weight  &  $\mathbf{x}_{i}$, $\mathbf{v}_{i}$, $f_{i}$  & The position, velocity, and objective function value of the $i$th agent, respectively  \\  \hline 
$\mathbf{x}_{\mathrm{pbest}, i}$, $f_{\mathrm{pbest}, i}$ & The position and objective function value of the personal optima of the $i$th agent, respectively & 
$\mathbf{x}_{\mathrm{gbest}}$, $f_{\mathrm{gbest}}$ & The position and objective function value of the global optima  \\ \hline
$\mathbf{z}_{\mathrm{s}}$ & The noise sampled from normal distribution  &   $\lambda$   & The carrier wavelength  \\ \hline
$(\theta_r,\phi_r)$  & The direction of a receiver &  $\eta$      & The efficiency of a VAA  \\ \hline
$\eta^{\mathrm{LoS}}$/$\eta^{\mathrm{NLoS}}$ & The attenuation factors for LoS/NLoS links  &  $\sigma^2$     & The noise power   \\ \hline
\end{tabular}
\end{table*}

\subsection{Routing Design and UAV Network Optimization}
\label{Routing Design and Network Optimization}

\noindent Traffic routing design for post-disaster UAV networks has been widely investigated in existing works. For example, Zhang \emph{et al.}~\cite{Zhang2023Qlearning} investigated the integration of cognitive radio with UAV swarms for emergency communications, and proposed a Q-learning-based architecture to obtain the intelligent routing. Sharvari \emph{et al.}~\cite{SharvariN2023} proposed an opportunistic routing scheme considering UAV coverage and collision constraints, to maximize the expected progress of each hop data packet. Moreover, Yang \emph{et al.}~\cite{Yang2023sensorjournal} constructed a three-layer UAV network to assist post-disaster relief and developed a greedy perimeter comprehensive evaluation routing algorithm. This algorithm selects the optimal next-hop UAV by assessing the link stability, throughput, and available energy of neighboring nodes. However, the works above do not consider UAV network optimization, potentially compromising transmission efficiency and reliability.

\par Several studies have investigated UAV network optimization for post-disaster communications. For example, He \emph{et al.}~\cite{he2021multi} considered a multi-hop task offloading framework in which multiple UAVs collaborate to offload task while simultaneously performing edge computing operations, and they proposed two algorithms to optimize the resource allocation and deployment of the UAVs. Rahmati \emph{et al.}~\cite{Rahmati2022dynamic} deployed multiple relaying UAVs to enhance the data rate between a base station and a user. Then, they maximized data flow of the constructed network by optimize UAV trajectories and transmit power. Lin \emph{et al.}~\cite{Lin2022TVT} formulated a maximum delay minimization problem for a post-disaster UAV network, and proposed an SDS-BSUM algorithm to optimize the transmit powers of all users. Moreover, Luan \emph{et al.}~\cite{luan2021hierarchical} constructed a UAV-enabled MEC emergency network. To minimize the average subtask completion time, they used game theory to reconstruct the network topology and proposed a scheduling mechanism for subtask management. However, the absence of optimal traffic routing design in these works may result in data congestion and degraded network performance.

\subsection{Solution Mechanisms}
\label{subsec: Solution Mechanisms}

\noindent Convex optimization is a widely used technique for solving optimization problems in UAV communications and networking. For example, Hu \emph{et al.}~\cite{hu2021uplink} deployed a UAV as a relay to facilitate uplink communication between a group of disconnected APs and a remote BS. Then, they formulated an uplink throughput maximization problem and adopted successive convex approximation (SCA) to handle the corresponding subproblems by introducing auxiliary variables. Niu \emph{et al.}~\cite{niu2022task} considered a distributed system consisting of multiple UAVs and mobile devices, both responsible for executing tasks offloaded from the control center, and proposed a convex optimization-based decision algorithm to tackle the task scheduling problem. Zhang \emph{et al.}~\cite{zhang2019trajectory} formulated a capacity maximization problem within a multi-UAV-enabled emergency communication network. Then, they decomposed the non-convex problem into two subproblems and used SCA as well as block coordinate update algorithm to handle them. However, not all optimization problems are suitable for solving with convex optimization. For example, some practical problems, such as UAV path planning, are non-convex. Transforming these non-convex problems into convex ones often involves complex mathematical transformations, which may not always be feasible or efficient in practice. Moreover, convex optimization cannot deal with complex and large-scale optimization problems.

\par Deep reinforcement learning (DRL) technique has also been widely used for handling optimization problems in disaster scenario. For example, Guan \emph{et al.}~\cite{guan2023cooperative} deployed multiple UAVs as relays to bridge links between mobile users in disaster area and BSs. They proposed a trajectory design mechanism based on the K-means strategy and multi-agent proximal policy optimization algorithms to optimize overhead and improve deployment efficiency. Wan \emph{et al.}~\cite{wan2024deep} dispatched UAVs for disaster data collection considering the time-varying data value, and proposed an attention-based DRL scheme for multi-UAV scheduling. Moreover, Zhang \emph{et al.}~\cite{zhang2021trajectory} deployed a UAV-mounted BS to collect data from ground users located in post-disaster area, and formulated a UAV trajectory design problem considering the available energy of ground users and obstacles. Specifically, they modeled the UAV as an agent, transformed the formulated problem into a constrained Markov decision-making process, and proposed a deep-Q-network-based algorithm to address the problem. However, rapid response is critical during post-disaster relief, especially in dynamic environments. The relatively long training time of DRL models may hinder their ability to provide timely responses.

\par In contrast to the works above, we propose a UAV-swarm enabled collaborative self-organizing network to establish links between a ground device in a post-disaster area and a remote AP. Then, we design the optimal traffic routing through theoretical analysis and optimize the communication network by using a heuristic-based algorithm, which  enables efficient and robust data transmission in post-disaster areas.

%
%
\section{System Model and Problem Formulation}
\label{sec: System Model and Problem Formulation}

\noindent In this section, we present the architecture of the UAV-swarm enabled collaborative self-organizing network. Then, we introduce the communication model which contains VAA model, channel model, and transmission model. Finally, we formulate the optimization problem. Note that vectors and matrices are represented by bold lowercase and uppercase letters, respectively, and Table \ref{table: notations} summarizes the notations in this paper.

\subsection{Network Architecture}
\label{subsec: Network Architecture}

\begin{figure}
\centerline{\includegraphics[width=3.5in]{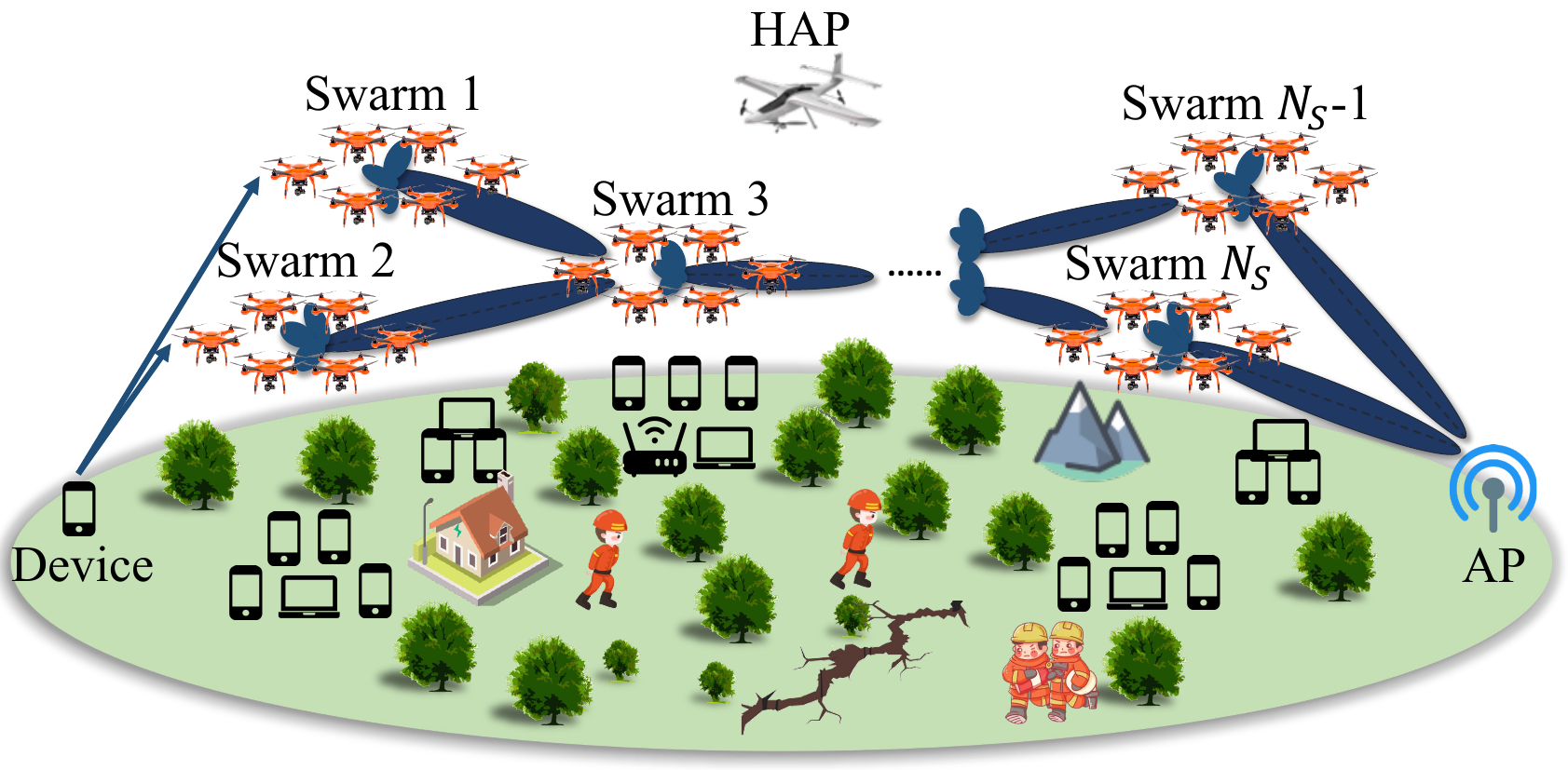}}
\caption{A UAV-swarm enabled collaborative self-organizing network for post-disaster communications.}
\label{fig. sketch map}
\end{figure}

\noindent The architecture of the proposed UAV-swarm enabled collaborative self-organizing network for post-disaster communications is illustrated in Fig. \ref{fig. sketch map}. Specifically, it primarily consists of the following components:

\begin{itemize}

\item A ground device in the post-disaster area which urgently needs to establish a communication link with external equipments to send distress signals.

\item A remote AP located outside the post-disaster area. Specifically, the AP enables the trapped devices to transmit urgent information and restore communication links with the external areas.

\item A UAV network comprising multiple UAV swarms, denoted as $\mathcal{U} = \{\mathcal{S}_{1}, \mathcal{S}_{2}, ..., \mathcal{S}_{N_{\mathrm{S}}}\}$. These UAV swarms are deployed over different spots of the post-disaster area to execute disaster missions, such as relaying data between devices and the remote AP. Each UAV swarm, denoted as $\mathcal{S}_{s} = \{1, 2, ..., N_{\mathrm{U}}\}$, is controlled by a ground rescue team for ease of management.
    
\item A high-altitude platform (HAP), serving as an essential coordination hub in LAE, which is responsible for executing algorithms and generating control strategies based on its powerful computational capabilities. These strategies are used to adjust system parameters such as positions of UAVs, to enhance system performance. Note that the HAP does not participate in data transmission.

\end{itemize}

\par
\begin{figure}
\centerline{\includegraphics[width=3.3in]{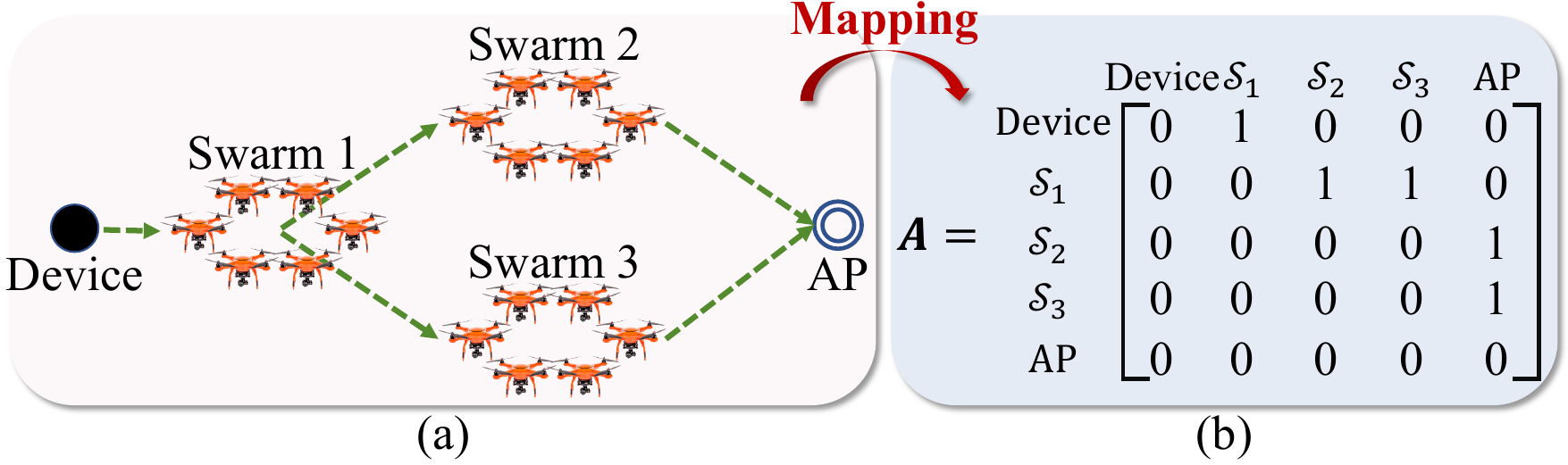}}
\caption{An example illustrating the mapping of the association relationship among components in the network and the adjacency matrix. (a) A two-dimensional sketch map of a simple self-organizing network on the ground. (b) The adjacency matrix of the self-organizing network.}
\label{fig: adjacency matrix}
\end{figure}

 Due to long distances, obstacles, not any pairs of the ground device, multiple UAV swarms, and the AP can establish a communication link. In this case, we introduce the generalized adjacency matrix $\mathbf{A}$ to record the connectivity relationships among the communication components. Specifically, the mapping of a UAV network and its adjacency matrix is shown in Fig. \ref{fig: adjacency matrix}. Accordingly, the working mechanism of the network for data transmission is described as follows.

\begin{enumerate}[label=\alph*)]

\item The ground device searches $\mathbf{A}$ and transmits data to all neighbors with which it can establish a link, by using a time division multiple access (TDMA) method.

\item If the receiver is the AP, the data transmission will stop. Otherwise, if a UAV swarm receives the data, it will search $\mathbf{A}$ and forward the received data to its neighbors by using the TDMA approach.

\item The transmission process will terminate once upon the AP receives all the intended data from different links.

\end{enumerate}

\par During the data forwarding process, it is assumed that a single UAV within the swarm receives the data and forwards it to other UAVs in the swarm for transmission.

\par We consider a three-dimensional (3D) Cartesian coordinate system, where the locations of the ground device, the AP, and the $u$th UAV in $\mathcal{S}_{s}$ are $\textbf{q}_{\mathrm{d}} = [x_{\mathrm{d}}, y_{\mathrm{d}}, z_{\mathrm{d}}]$, $\textbf{q}_{\mathrm{a}} = [x_{\mathrm{a}}, y_{\mathrm{a}}, z_{\mathrm{a}}]$, and $\textbf{q}_{u}^{s} = [x_{u}^{s}, y_{u}^{s}, z_{u}^{s}]$, respectively. Note that the locations of the ground device and the AP are static, while the movement area of UAVs in $\mathcal{S}_{s}$ is confined in $\mathbb{R}_s^3$ for ease of control.

\subsection{Communication Model}
\label{subsec: Communication Model}

\subsubsection{Virtual Antenna Array Model}
\label{subsubsec: VAA Model}

\noindent The array factor (AF) is used to characterize the beam pattern of each VAA. Specifically, the AF of the VAA formed by UAVs in the $s$th UAV swarm is given by~\cite{Li2021}:
\begin{equation}
\small
\label{eq. AF}
F_s(\theta,\phi)=\sum\nolimits_{u=1}^{N_{\mathrm{U}}}I_u^{s}\text{e}^{i_{\mathrm{u}}[p(x_{u}^{s}\sin\theta\cos\phi + y_{u}^{s}\sin\theta\sin\phi + z_{u}^{s}\cos\theta)]},
\end{equation}
\noindent where $i_{\mathrm{u}}$ represents the imaginary unit, and $p = 2\pi/\lambda$ denotes the phase constant with $\lambda$ being the carrier wavelength.

\par Accordingly, through regarding the whole antenna array as a single directional antenna, the antenna gain $G_{s}$ of the VAA formed by UAVs in the $s$th UAV swarm is given by~\cite{Mozaffari2019LAA}:
\begin{equation}
\label{eq. G}
G_s = \frac{4\pi|F_s(\theta_r,\phi_r)|^2w(\theta_r,\phi_r)^2}{\int_0^{2\pi}\int_0^{\pi}|F_s(\theta,\phi)|^2w(\theta,\phi)^2\sin\theta\,d\theta\,d\phi}\eta,
\end{equation}
\noindent where $(\theta_r,\phi_r)$ refers to the direction of the receiver. Moreover, $w(\theta,\phi)$ is the magnitude of the far-field beam pattern of each UAV element, and $\eta$ denotes the antenna array efficiency.

\subsubsection{Channel Model}
\label{subsubsec: Channel Model}

\noindent The network comprises three types of communication links, which are ground-to-air (G2A) links, air-to-air (A2A) links, and air-to-ground (A2G) links, respectively. In this work, a probabilistic LoS model is used to characterize these communication links. Note that the parameters of the A2A links are different from those of the other two links. Specifically, the path loss between nodes $i$ and $j$ is as follows~\cite{Rahmati2022interference}:
\begin{equation}
PL_{i,j} = \left\{\begin{array}{ll}
(K_{\mathrm{o}} d_{i,j})^{\alpha}\eta^{\mathrm{LoS}}, & \text{LoS link}\\
(K_{\mathrm{o}} d_{i,j})^{\alpha}\eta^{\mathrm{NLoS}}, & \text{non-LoS link},
\end{array}\right.
\label{eq. los and nlos path loss}
\end{equation}
\noindent where $K_{\mathrm{o}} = 4\pi f_\mathrm{c}/c$ is the path loss constant, $f_\mathrm{c}$ refers to the carrier frequency and $c$ is the speed of light. Moreover, $d_{i,j}$ is the total distance between nodes $i$ and $j$, and $\alpha$ is the path loss exponent. In addition, $\eta^{\mathrm{LoS}}$ and $\eta^{\mathrm{NLoS}}$ are attenuation factors for LoS and non-NLoS (NLoS) links, respectively.

\par The probability that a LoS link exists between nodes $i$ and $j$ is as follows~\cite{AlHourani2014}:
\begin{equation}
\mathbb{P}_{i,j}^{\mathrm{LoS}} = \frac{1}{1 + m \, \text{exp}\left[-n\left(180/\pi\arcsin(H_{i, j}/d_{i, j}) - m\right)\right]},
\label{eq. LoS environment}
\end{equation}
\noindent where $m$ and $n$ are constants that depend on environment. Moreover, $H_{i, j}$ represents the vertical distance between nodes $i$ and $j$. Accordingly, the probability that a NLoS link exists between nodes $i$ and $j$ is $\mathbb{P}_{i,j}^{\mathrm{NLoS}} = 1- \mathbb{P}_{i,j}^{\mathrm{LoS}}$.

\par Based on this, the channel power gain between nodes $i$ and $j$ is given by:
\begin{equation}
h_{i, j} = (K_{\mathrm{o}} d_{i,j})^{-\alpha}[\mathbb{P}^{\mathrm{LoS}}_{i,j}\eta^{\mathrm{LoS}} + \mathbb{P}^{\mathrm{NLoS}}_{i,j}\eta^{\mathrm{NLoS}}]^{-1}.
\label{eq. channel power gain between nodes i and j}
\end{equation}

\subsubsection{Transmission Model}
\label{subsubsec: Transmission Model}

\noindent Given the VAA and channel models described above, the transmission rate $R_{i,j}$ between nodes $i$ and $j$ is given by~\cite{Zeng2017efficiency}:
\begin{equation}
R_{i, j} = B\log_2\left(1+P_{i}G_{i}h_{i,j}/\sigma^2\right),
\label{eq. transmission rate between two nodes}
\end{equation}
\noindent where $B$ and $\sigma^2$ refer to the channel bandwidth and noise power, respectively. Moreover, $P$ denotes the transmit power of node $i$. If node $i$ is the ground device, then $P = p_t$, where $p_{\mathrm{t}}$ denotes the transmit power of the ground device or a single UAV. Conversely, if node $i$ represents the $s$th UAV swarm, then $P_{i}$ is given by $P_{i} = \sum\nolimits_{u=1}^{N_{\mathrm{U}}}(I_{u}^{s})^2p_{\mathrm{t}}$. In addition, $G_{i}$ is the antenna gain of node $i$, which is as follows:
\begin{equation}
G_{i} = \left\{\begin{array}{ll}
1, & \text{node $i$ is the ground device}\\
G_s, & \text{node $i$ is the $s$th UAV swarm}.
\end{array}\right.
\label{eq. antenna gain in transmission model}
\end{equation}

\par Note that UAV coordination within the same swarm before data transmission is assumed to be handled by an existing algorithm~\cite{lamport2001paxos} and is beyond the scope of this work. 

\subsection{Energy Consumption Model}
\label{sec: Energy Consumption Model}

\noindent When the $u$th UAV in the $s$th swarm flies in a 2D plane with speed $v_{u}$, the propulsion power consumption is as follows:
\begin{align}
P(v_{u})= & P_{\mathrm{0}}\left(1+\frac{3 v_{u}^{2}}{v_{\mathrm{b}}^{2}}\right) + P_{\mathrm{I}} \left(\sqrt{1 + \frac{v_{u}^{4}}{4v_{\mathrm{m}}^{4}}}-\frac{v_{u}^{2}}{2v_{\mathrm{m}}^{2}} \right)^{1/2} \nonumber  \\ 
& + \frac{1}{2} d_0 \rho sAv_{u}^3, \label{eq. propulsion power cons of UAVs in 2D}
\end{align}
\noindent where $P_{\mathrm{0}}$ and $P_{\mathrm{I}}$ are the blade profile power and induced power in hovering status, respectively. Moreover, $v_{\mathrm{b}}$ and $v_{\mathrm{m}}$ are the tip speed of the rotor blade and the mean rotor induced velocity in hovering, respectively. In addition, $d_0$, $\rho$, $s$, and $A$ are the fuselage drag ratio, air density, rotor solidity, and rotor disc area, respectively. Accordingly, for the $u$th UAV in the $s$th swarm with a 3D trajectory $\mathbf{q}_{u}(t)$, its energy consumption is given by~\cite{Zeng2019accessing}:
\begin{align}
E_{u}^{s}(\mathbf{q}^{s}_{u}(t)) \approx & \int_{0}^{T}P(v_{u}(t))dt + \frac{1}{2} m_{u}\left(v_{u}(T)^2 - v_{u}(0)^2\right) \nonumber \\
& + m_{u}g\left(h(T)-h(0)\right), \label{eq: propulsion power cons of UAVs in 3D}
\end{align}
\noindent where $m_{u}$ and $g$ represent the mass of the $u$th UAV and gravitational acceleration, respectively. Moreover, $T$ is the flight time of the UAV. Accordingly, the flight energy consumption of the whole system is $E = \sum\nolimits_{s = 1}^{N_{\mathrm{S}}}\sum\nolimits_{u = 1}^{N_{\mathrm{U}}} E_{u}^{s}$.

\subsection{Problem Formulation}
\label{subsec: Problem Formulation}

\noindent In post-disaster areas, critical data such as rescue requests, health information, and situational reports must be delivered to remote base stations or APs timely. Moreover, communication opportunities may be brief in post-disaster areas due to the environmental challenges and the mobility of UAVs, etc. In this case, to guarantee that the urgent data can be relayed promptly and reduce the risk of data loss, we formulate an RPTRMOP aimed at maximizing the data transmission rate $R_{d, a}$ between the ground device and the AP. Accordingly, the RPTRMOP is given by:
\begin{subequations}
\label{eq. RPTRMOP}
\begin{align}
\underset{\mathbf{I}, \mathbf{P}}{\text{max}} \ &  R_{d, a},  \label{subeq. RPTRMOP} \\
\textup{s.t.}\
&C1: 0 \le I_{u}^s \le 1, \forall s\in \mathcal{U}, u\in \mathcal{S}_{s}, \label{subeq. RPTRMOP C1}\\
&C2: [x_{u}^{s}, y_{u}^{s}, z_{u}^{s}] \in \mathbb{R}_s^3, \forall s\in \mathcal{U}, u\in \mathcal{S}_{s}, \label{subeq. RPTRMOP C2}\\
&C3: D_{u_1,u_2} \ge (D_{\mathrm{min}},\lambda/2), \forall u_1, u_2 \in \mathcal{S}_{s}, \label{subeq. RPTRMOP C3}\\
&C4: E \leq E_{\mathrm{th}}, \label{subeq. RPTRMOP C4 energy constraint}
\end{align}
\end{subequations}
\noindent where $\mathbf{I}$ and $\mathbf{P}$ are decision variables of the formulated RPTRMOP, with $\mathbf{I}$ denoting the set of UAV excitation current weights, which is directly associated with the transmit power and $\mathbf{P}$ being the set of 3D placements of UAVs, respectively. Moreover, the constraint \eqref{subeq. RPTRMOP C1} restricts the value of the excitation current weight of each UAV to [0, 1], the constraint \eqref{subeq. RPTRMOP C2} confines the 3D placements of UAVs in the $s$th swarm to a predefined area $\mathbb{R}_s^3$, and the constraint \eqref{subeq. RPTRMOP C3} ensures that the distance between two UAVs in the $s$th swarm exceeds a specified threshold to avoid collisions. In addition, the constraint \eqref{subeq. RPTRMOP C4 energy constraint} ensures the flight energy consumption of the entire system lower than the threshold $E_{\mathrm{th}}$.

\par Though the structure is seemingly simple, the formulated RPTRMOP is intractable. The reason is that $R_{d, a}$ cannot be expressed with a specific formula. \emph{On the one hand}, optimizing $R_{d, a}$ requires first determining the optimal traffic routing. However, the considered network has a mesh structure, which implies that the topologies among UAV swarms are complex, making the determination of the traffic routing for data transmission particularly challenging. \emph{On the other hand}, even after the traffic routing is determined, the excitation current weights and positions of the UAVs within each swarm involved in data transmission must be optimized to achieve the optimal transmission rate for each communication link. Then, the optimal transmission rate between the ground device and the AP can be attained.

\par Given the aforementioned challenges, we propose a two-stage optimization approach to solve the formulated problem. In the first stage, the optimal traffic routing is obtained via theoretical analysis and an existing method. In the second stage, the network is optimized based on the routing derived from the first stage.

\section{Stage One: Routing Design}
\label{sec: Stage One}

\noindent In this section, we first derive the theoretical upper bound on the transmission rate of each link in the communication network. Then, we model the network as a directed weighted graph, and classify the formulated RPTRMOP as a maximum flow problem. Finally, we deduce the optimal traffic routing and the theoretical upper bound on the transmission rate of the entire network by using an existing method.

\subsection{Derivation of Theoretical Upper Bound on the Transmission Rate of Each Communication Link}
\label{subsec: Deduction of rate upper bound of each communication link}

\noindent In this part, we show the theoretical upper bound on the transmission rate of each communication link in the network. Specifically, we first deduce the theoretical upper bound on the signal-to-noise ratio (SNR) of each communication link in the following lemma.

\begin{Lemma} 
\label{Lemma: maximum gain of SNR}
Consider a collaborative system comprising $N_{\mathrm{U}}$ UAVs that form a VAA. Let $\rho$ denote the SNR achieved by a single UAV transmission. Then, the maximum achievable SNR of the VAA system, denoted as $\rho_{N_{\mathrm{U}}}$, is upper-bounded by:
\begin{equation}
\rho_{N_{\mathrm{U}}} \leq N_{\mathrm{U}}^2\rho,
\end{equation}
where the equality holds under ideal beamforming conditions.
\end{Lemma}
\emph{Proof:} The proof follows from two key aspects. \emph{First}, let $p_{\text{t}}$ denote the transmit power of a single UAV. When all UAVs operate at maximum excitation current weights, the total transmit power of a VAA $P_{\text{total}}$ satisfies $ P_{\text{total}} = \sum_{u=1}^{N_{\mathrm{U}}} p_{\text{t}} = N_{\mathrm{U}}p_{\text{t}}$. \emph{Second}, under ideal beamforming conditions, the array gain provides an additional factor of $N_{\mathrm{U}}$ in received power due to coherent signals combining~\cite{Dong2008N2SNR}. Consequently, the maximum achievable SNR of the VAA system is $\rho_{N_{\mathrm{U}}} = N_{\mathrm{U}} \cdot N_{\mathrm{U}}\rho = N_{\mathrm{U}}^2\rho$.
{\hfill $\blacksquare$\par}

\par Based on Lemma \ref{Lemma: maximum gain of SNR}, the theoretical upper bound on the transmission rate of each communication link in the network can be acquired in the following theorem.

\begin{Theorem}
\label{theorem: theoretical transmission rate upper bound}
For a collaborative system with $N_{\mathrm{U}}$ UAVs, let $B$ denote the channel bandwidth, $h$ the channel power gain, $p_{\mathrm{t}}$ the transmit power of each UAV, and $\sigma^2$ the noise power. Then, the theoretical upper bound on the achievable transmission rate $C$ is given by:
\begin{equation}
C \leq B\log_2\left(1 + \frac{N_{\mathrm{U}}^2p_{\mathrm{t}}h}{\sigma^2}\right).
\end{equation}
\end{Theorem}
\emph{Proof:} For a single UAV transmit data with power $p_{\mathrm{t}}$, the achieved SNR is $\rho = p_{\mathrm{t}}h/\sigma^2$. Based on Lemma \ref{Lemma: maximum gain of SNR}, the achieved SNR of a VAA system with $N_{\mathrm{U}}$ UAVs is $\rho_{N_{\mathrm{U}}} = N_{\mathrm{U}}^2p_{\mathrm{t}}h/\sigma^2$ under ideal beamforming conditions. Thus, by applying the Shannon-Hartley theorem, the theoretical upper bound on the transmission rate of the VAA system is $C = B\log_2\left(1 + N_{\mathrm{U}}^2p_{\mathrm{t}}h/\sigma^2\right)$.
{\hfill $\blacksquare$\par}

\par We obtain the theoretical upper bound on the transmission rate of a VAA based on Lemma \ref{Lemma: maximum gain of SNR} and Theorem \ref{theorem: theoretical transmission rate upper bound}. Subsequently, the optimal traffic routing and the theoretical upper bound on the transmission rate of the network are obtained.

\subsection{Traffic Routing Design}
\label{subsec: Routing Design}

\noindent In this section, we first model the communication network as a directed weighted graph. Then, the formulated RPTRMOP is classified as a maximum flow problem. Finally, we obtain the optimal traffic routing and the theoretical upper bound on the transmission rate of the network by using Ford-Fulkerson algorithm.

\subsubsection{Graph Construction}
\label{subsubsec: Graph Construction}

\noindent The constructed network can be represented as a directed weighted graph $G = \{\mathcal{V}, \mathcal{E}, \mathcal{W}\}$, in which $\mathcal{V}$ denotes the set of all nodes in the network, $\mathcal{E}$ is the set of communication links among the nodes, and $\mathcal{W}$ specifies the capacity of each communication link in $\mathcal{V}$. Specifically, the graph is constructed as follows:

\begin{itemize}

\item [a)] \emph{Label Network Nodes}. Each UAV swarm is treated as a distinct entity and labeled sequentially as $1$, $2$, ..., $N_{\mathrm{S}}$. Moreover, the device and AP are represented by $0$ and $N_{\mathrm{S}}+1$, respectively. Therefore, the set of all nodes in the network, $\mathcal{V}$, is constructed as 
\begin{equation}
\label{eq. label network nodes}
\mathcal{V} = \{0, 1, 2, ..., N_{\mathrm{S}}+1\}.
\end{equation}
    
\item [b)] \emph{Define the Connectivity among Nodes}. The adjacency matrix $\mathbf{A}$ is used to represent the connectivity among the nodes in $\mathcal{V}$. For each pair of nodes $i$ and $j$ in $\mathcal{V}$, if $a_{i,j} = 1$ (i.e., there is a link between nodes $i$ and $j$), then $<i,j>$ is placed into the set of communication links $\mathcal{E}$ as follows:
\begin{equation}
\mathcal{E} \gets <i,j>, \{i,j\} \in \mathcal{V}, i \neq j.
\label{eq. construct edge}
\end{equation}
\item [c)] \emph{Calculate the Capacity of Each Communication Link}. For each $<i,j> \in \mathcal{E}$, calculate the maximum capacity $C_{i,j}$ based on Theorem \ref{theorem: theoretical transmission rate upper bound}, which provides the theoretical upper bound on the transmission rate between nodes $i$ and $j$. Then, $C_{i,j}$ is put into the set of capacities $\mathcal{W}$ as follows:
\begin{equation}
\mathcal{W} \gets C_{i,j}, <i, j> \in \mathcal{E}, i \neq j.
\label{eq. construct edge}
\end{equation}
\end{itemize}

\subsubsection{Problem Reformulation}
\label{subsubsec: Problem Reformulation}

\noindent Given the graph $G$ above, the intractable problem formulated in Eq. \eqref{eq. RPTRMOP} can be reformulated as follows:
\begin{subequations}
\label{eq. reformulated problem}
\begin{align}
\underset{\mathbf{I}, \mathbf{P}}{\text{max}} \ & R_{0, N_{\mathrm{S}}+1},  \label{subeq. reformulated RPTRMOP} \\
\textup{s.t.}\  
& C1: \sum\limits_{<i,j>\in \mathcal{E}} R_{i,j} - \sum\limits_{<j,k>\in \mathcal{E}} R_{j,k} = 0, \notag \\ 
& \quad \quad \forall j \in\mathcal{V} \backslash \{0,N_{\mathrm{S}}+1\}, \label{subeq. reformulated problem C1}\\
& C2: 0 \leq R_{i,j} \leq C_{i,j}, \forall <i,j>\in\mathcal{E}, \label{subeq. reformulated problem C2}\\
& \eqref{subeq. RPTRMOP C1}, \eqref{subeq. RPTRMOP C2}, \eqref{subeq. RPTRMOP C3}, \eqref{subeq. RPTRMOP C4 energy constraint},\notag
\end{align}
\end{subequations}
\noindent where the constraint \eqref{subeq. reformulated problem C1} shows the assumption of balanced transmission rates for all UAV swarms. Moreover, the constraint \eqref{subeq. reformulated problem C2} ensures that the transmission rate of each link in $\mathcal{E}$ does not exceed the derived maximum capacity. Based on the architecture of the reformulated problem, we have the following theorem.

\begin{Theorem}
The reformulated optimization problem in Eq. \eqref{eq. reformulated problem} is a typical maximum flow problem.
\label{Theorem: reformulated problem is a maximum flow problem}
\end{Theorem}
\emph{Proof:} The analysis is conducted from three key aspects. First, the network is modeled as a directed weighted graph $G$, consistent with the structure of a maximum flow problem. Moreover, our objective is to maximize the transmission rate $R_{0, N_{\mathrm{S}}+1}$ between the ground device and the AP, corresponding to the goal of maximizing flow in a typical maximum flow problem. Finally, the constraint \eqref{subeq. reformulated problem C1} ensures balanced inflow and outflow at the UAV swarms, while the constraint \eqref{subeq. reformulated problem C2} limits the transmission rate of each link to its capacity. Both constraints are key conditions in a maximum flow problem. Thus, the reformulated problem is a typical maximum flow problem~\cite{Rahmati2019INFOCOMflow}. 

\par Given the aforementioned property of the reformulated problem, the well-known Ford-Fulkerson algorithm is adopted to deal with it in the following.

\begin{algorithm}[htbp]
	\caption{Ford-Fulkerson Algorithm}
	\label{algo: Ford-Fulkerson Algorithm}
	\KwIn{The directed weighted graph $G = \{\mathcal{V}, \mathcal{E}, \mathcal{W}\}$.}
	\KwOut{The theoretical upper bound on the transmission rate of the network $\overline{R}_{0, N_{\mathrm{S}}+1}$ and a optimized network $G'$.}
    $\overline{R}_{0, N_{\mathrm{S}}+1} \gets 0$; \\
    \For{$<i,j> \in \mathcal{E}$}
	{
        $R_{i,j} = 0$;\\
        $R_{j,i} = 0$;
    }
    \While{\rm{there is a path} $p_{\mathrm{a}}$ in the residual network $G_r$}{
        $c_{r}(p_{\mathrm{a}})$ = min\{$c_{r}(i,j)$, $<i,j>$ $\in$ $p_{\mathrm{a}}$\};\\
        \For{\rm{each} $<i,j>$ $\in p_{\mathrm{a}}$}
        {
            $R_{i,j} = R_{i,j} + c_{r}(p_{\mathrm{a}})$;\\
            $R_{j,i} = -R_{i,j}$;\\
        }
        $\overline{R}_{0, N_{\mathrm{S}}+1} \gets \overline{R}_{0, N_{\mathrm{S}}+1} + c_{r}(p_{\mathrm{a}})$;
    }
    Return $\overline{R}_{0, N_{\mathrm{S}}+1}$, $G'$.
\end{algorithm}

\subsubsection{Ford-Fulkerson Algorithm}
\label{subsubsec: Ford-Fulkerson Algorithm}

\noindent Several existing methods can calculate the theoretical upper bound on the transmission rate $R_{0, N_{\mathrm{S}}+1}$ between the ground device and the AP~\cite{fordfulkersson2015flows}, such as the Ford-Fulkerson algorithm, the Edmonds-Karp algorithm, and the Dinic Push-Relabel algorithm. Due to its simplicity, efficiency, and applicability to any network, the Ford-Fulkerson algorithm is adopted in this study. To enhance understanding of the algorithm, several technical terms are explained first, as detailed in~\cite{fordfulkersson2015flows}.

\begin{Definition}
(Residual capacity): For a pair of nodes $i$ and $j$ in the network, the residual capacity of the link $<i,j>$ refers to the maximum additional transmission rate that can be pushed from $i$ to $j$ without exceeding the corresponding capacity $C_{i, j}$, which is defined as $c_{i,j} = C_{i, j} - R_{i, j}$.
\label{Definition: Residual capacity}
\end{Definition}

\begin{Definition}
(Reverse Link): Reverse link is a introduced virtual link. For a link $<i,j>$ with transmission rate $R_{i, j}$, a corresponding reverse link exists in the residual network from $j$ to $i$, with a residual capacity $c_{j,i} = R_{i, j}$.
\label{Definition: Reverse Edge}
\end{Definition}

\begin{Definition}
(Residual network): For an original network $G$, the residual network is $G_{\mathrm{r}} = \{\mathcal{V}, \mathcal{E}_{r}, \mathcal{W}_{r}\}$, where $\mathcal{E}_{r} = \mathcal{E} \cup \mathcal{E}_{a}$, with $\mathcal{E}_{a} = \{<j,i>|<i,j> \in \mathcal{E}, i \neq j\}$ being the set of reverse links, and $\mathcal{W}_{r}$ is the set of residual capacities of links in $\mathcal{E}_{r}$.
\label{Definition: Residual Network}
\end{Definition}

\begin{Definition}
(Augmenting path): Given an original network $G$, an augmenting path $p_{\mathrm{a}}$ is a path from 0 to $N_{\mathrm{S}}+1$ in the residual network $G_{\mathrm{r}}$. The introduction of augmenting path can increase the total transmission rate from $0$ to $N_{\mathrm{S}}+1$.
\label{Definition: Augmenting Path}
\end{Definition}

\par Based on the definitions provided above, the main framework of the Ford-Fulkerson algorithm for solving the reformulated problem in Eq. \eqref{eq. reformulated problem} is outlined in Algorithm \ref{algo: Ford-Fulkerson Algorithm}. The detailed steps are presented as follows:

\begin{itemize}

\item [a)] \emph{Initialization}. Set the transmission rate between 0 and $N_{\mathrm{S}}+1$ as 0. Moreover, set the transmission rate of each link $<i,j> \in \mathcal{E}$ in $G$ and its reverse link $<j,i>$ to 0.

\item [b)] \emph{Augmented Path Finding}. Check if there exists any augmenting path in the residual network $G_{\mathrm{r}}$. If a path $p_{\mathrm{a}}$ is found, define the residual capacity of the path $c_r(p_{\mathrm{a}})$ as the minimum residual capacity of all links in $p_{\mathrm{a}}$. 

\item [c)] \emph{Transmission Rate Update}. For each link $<i,j>$ in $p_{\mathrm{a}}$, add the calculated path capacity $c_r(p_{\mathrm{a}})$ to the transmission rate $R_{i,j}$. Simultaneously, adjust the transmission rate of the reverse link $R_{j,i}$ by setting it to the negative of the updated $R_{i,j}$. Then, increment the total transmission rate $R_{0, N_{\mathrm{S}}+1}$ by $c_{r}(p_{\mathrm{a}})$.

\item [d)] \emph{Termination}. Continue iterating through the second step until no further augmenting paths are found in $G_{\mathrm{r}}$.

\end{itemize}

\par After the algorithm is executed, the theoretical upper bound on the transmission rate of the network $\overline{R}_{0, N_{\mathrm{S}}+1}$ is obtained, accompanied by a optimized network $G'$. Specifically, $G'$ is obtained by eliminating the nodes which no not need to execute any data receiving or forwarding tasks at all, along with the links whose transmission rates are zero. Akin to the structure of the original network $G$, the components in the optimized network $G' = (\mathcal{V}', \mathcal{E}', \mathcal{W}')$ is as follows:

\begin{itemize}

\item \emph{$\mathcal{V'} = \{0, ..., N_{\mathrm{S}}', N_{\mathrm{S}}'+1\}$} contains the ground device, the $N_{\mathrm{S}}'$ UAV swarms that participate in the data transmission, and the AP, subject to $\mathcal{V}' \subseteq \mathcal{V}$.

\item \emph{$\mathcal{E'} = \{<i,j>|R_{i,j} \neq 0 \ \text{or} \ R_{j,i} \neq 0\}$} refers to the set of links among nodes in $\mathcal{V'}$.

\item \emph{$\mathcal{W'} = \{\overline{R}_{i,j}| <i,j> \in \mathcal{E'}\}$} represents the set of expected transmission rate of each link in $\mathcal{E'}$.
		
\end{itemize}

\par
\begin{figure}
\centerline{\includegraphics[width=3.5in]{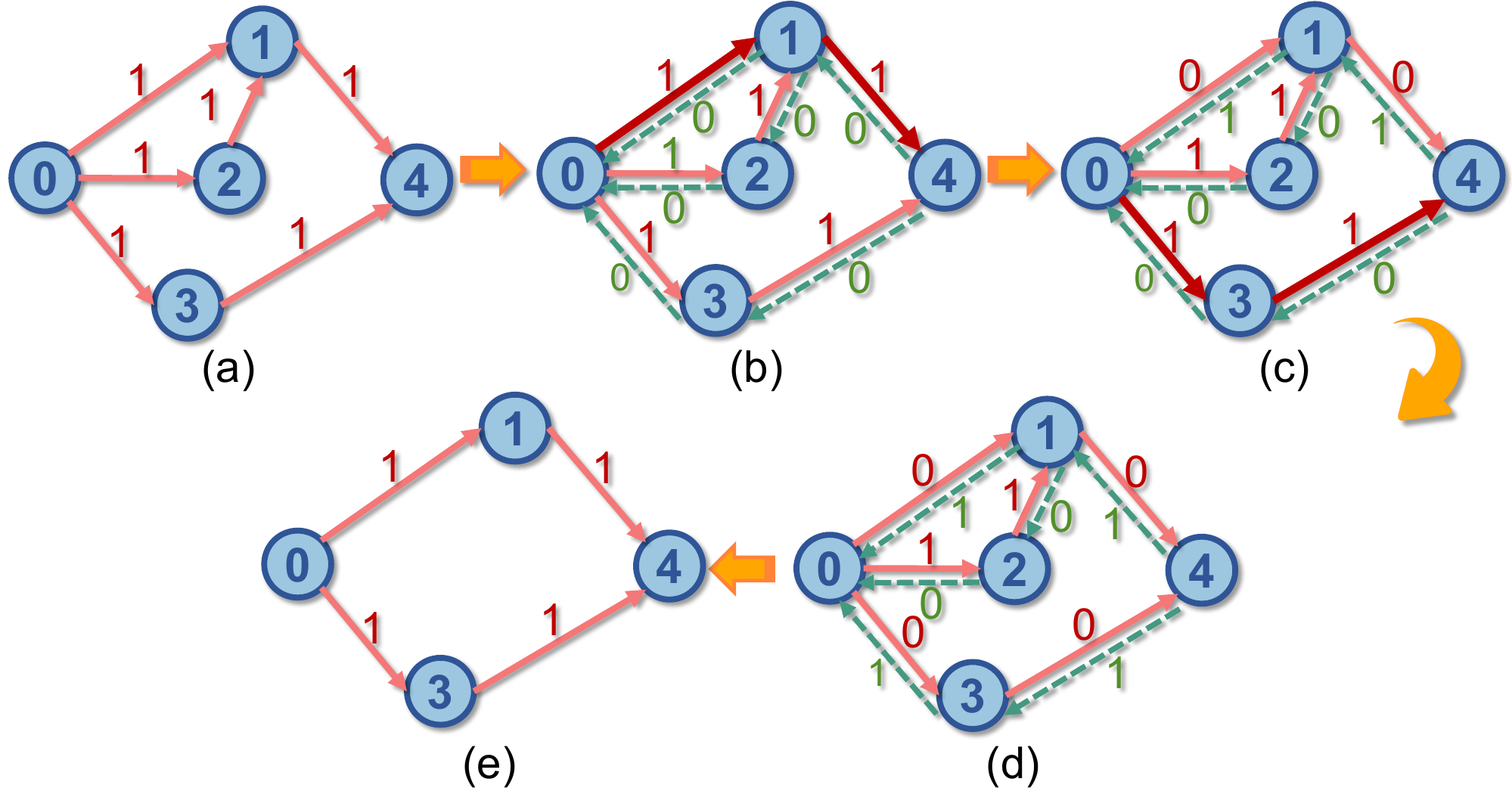}}
\caption{An example illustrating the Ford-Fulkerson algorithms for calculating the theoretical upper bound on the transmission rate of a network and the optimal traffic routing. (a) The original network, where node 0 is the transmitter and the node 4 is the receiver. (b) A residual network of (a), where an augmenting path 0$\rightarrow$1$\rightarrow$4 exists, thus the current upper bound on the transmission rate is 1. (c) A residual network of (a), where an augmenting path 0$\rightarrow$3$\rightarrow$4 exists, thus the current upper bound on the transmission rate is 2. (d) A residual network of (a) with no augmenting paths. (e) The final optimized network with an upper bound on the transmission rate of 2 and two corresponding paths.}
\label{fig. process of obtaining the optimized network}
\end{figure}

 Intuitively, $G'$ provides the optimal traffic routing with several simple paths for parallel data transmission~\cite{jeon2013fully}. Moreover, the transmission rate for each link in each path corresponds to its optimal value. Fig. \ref{fig. process of obtaining the optimized network} provides a simple example illustrating the Ford-Fulkerson algorithm for calculating the theoretical upper bound on the transmission rate of a network, and demonstrates the relationships among the original, residual, and final optimized networks.

\par The computational complexity of the Ford-Fulkerson algorithm in our optimization problem is analyzed in the following theorem. Notably, since the Ford-Fulkerson algorithm guarantees the optimal maximum flow solution, the complexity analysis represents both the best case and worst case scenarios.

\begin{Theorem}
\label{Theorem: complexity analysis of Ford-Fulkerson}
Given that the number of links in the set $\mathcal{E}$ is $N_{L}$ and the theoretical upper bound on the transmission rate of the system is $\overline{R}_{0, N_{\mathrm{S}}+1}$, the Ford-Fulkerson algorithm achieves a computational complexity of $\mathcal{O}\left(N_{L}\overline{R}_{0, N_{\mathrm{S}}+1}\right)$ when solving the maximum flow problem formulated in Eq.~\eqref{eq. reformulated problem}.
\end{Theorem}
\emph{Proof:} The complexity analysis is derived from the operation of the Ford-Fulkerson algorithm in this problem setting. The algorithm identifies augmenting paths iteratively using depth-first searches, where each path increases the flow value by at least one unit. Since the theoretical upper bound on the transmission rate is bounded by $\overline{R}_{0, N_{\mathrm{S}}+1}$, the algorithm requires at most $\overline{R}_{0, N_{\mathrm{S}}+1}$ iterations to identify all augmenting paths. In each iteration, a depth-first search may traverse all $N_{L}$ links in the set $\mathcal{E}$, resulting in a per-iteration complexity of $\mathcal{O}(N_{L})$. Therefore, the total computational complexity is given by $\mathcal{O}\left(N_{L}\overline{R}_{0, N_{\mathrm{S}}+1}\right)$.
{\hfill $\blacksquare$\par}

\par Based on the theoretical analysis of the first stage, we obtain the optimal traffic routing and the theoretical upper bound on the transmission rate of the network. In the following, the excitation current weight and placement of each participating UAV are designed to optimize the communication network, ensuring that the actual transmission rate approaches its theoretical upper bound, i.e., $\overline{R}_{0, N_{\mathrm{S}}+1}$.

\section{Stage Two: UAV Swarm Optimization}
\label{sec: Step Two}

\noindent In this section, we first transform the formulated RPTRMOP into its penalty-based variant, V-RPTRMOP, and provide a detailed analysis of the V-RPTRMOP. Then, we propose an optimization method to solve the V-RPTRMOP effectively.

\subsection{Problem Transformation and Analysis}
\label{subsec: problem transformation}

\noindent In this work, four Parts are considered when transforming the formulated RPTRMOP into a more tractable format, which are as follows:

\begin{itemize}

\item \emph{\textbf{Motivation I: Diminishing the gap between the actual and expected transmission rates of each communication link.}} To render the communication network closely align with the optimized network, the transmission rate $R_{i, j}$ of the link $<i,j>$ should be closely approach its theoretical upper bound, i.e., $\overline{R}_{i, j}$. Accordingly, the sum of differences between the actual and expected transmission rates of all links should be minimized.

\item \emph{\textbf{Motivation II: Mitigating extreme cases.}} If only proposal I is considered, extreme situations may occur, where the differences between the actual and expected transmission rates for different links become excessively huge. For instance, with $\overline{R}_{1,2} = 1.0 \times 10^7$ bps and $\overline{R}_{1,3} = 1.2 \times 10^7$ bps, the result could lead to $R_{1,2} = 1 \times 10^6$ bps and $R_{1,3} = 1.2 \times 10^7$ bps. In this case, only the second link achieves an acceptable transmission rate, while the counterpart of the other link is not optimized at all. To avoid such extreme cases, standard deviation is introduced as a corrective measure.

\item \emph{\textbf{Motivation III: Preventing communication network congestion.}} In the constructed communication network, UAV swarms are responsible for both receiving and transmitting data. If the actual transmission rate $R_{i,j}$ falls below its theoretical upper bound $\overline{R}_{i,j}$, the achievable rate of the receiver $j$ will be adversely affected, since it will not be able to transmit sufficient data, potentially leading to network congestion. To address this issue, a penalty mechanism is implemented. Specifically, the penalty term $\delta_{i,j}$ of the link $<i, j> \in \mathcal{E}$ is given by:
\begin{equation}
\delta_{i,j}=\left\{\begin{array}{ll}
1, & \text{if} \, R_{i,j} < \overline{R}_{i,j} \\
0, & \text{if} \, R_{i,j} \geq \overline{R}_{i,j}.
\end{array}\right.
\label{eq. penalty term}
\end{equation}

\item \emph{\textbf{Motivation IV: Avoiding excessive flight energy consumption.}} Excessive energy consumption can significantly diminish the endurance of UAVs, thereby restricting their flight range and mission duration. Thus, an additional penalty term $\eta$ is introduced to regulate flight energy consumption and extend UAV operational time, which is given by:
\begin{equation}
\eta=\left\{\begin{array}{ll}
1, & \text{if} \, E > E_{\mathrm{th}} \\
0, & \text{if} \, E \leq E_{\mathrm{th}}.
\end{array}\right.
\label{eq. penalty term}
\end{equation}
\end{itemize}

\par Based on the aforementioned considerations, the V-RPTRMOP is as follows:
\begin{subequations}
\label{eq. V-RPTRMOP}
\begin{align}
\underset{\mathbf{I}, \mathbf{P}}{\text{min}} \ f = & \underbrace{\alpha_{1}\sum_{\langle i, j \rangle \in \mathcal{E}}|R_{i,j} - \overline{R}_{i,j}|}_\text{Part I} \notag + \underbrace{\alpha_{2}\sqrt{\sum_{\langle i, j \rangle \in \mathcal{E}}\frac{|R_{i,j} - \overline{R}_{i,j}|^{2}}{N_{\mathrm{L}}'}}}_\text{Part II} \notag \\
& + \underbrace{\alpha_{3}\sum_{\langle i, j \rangle \in \mathcal{E}}\delta_{i,j}}_\text{Part III} + \underbrace{\alpha_{4}\eta}_\text{Part IV}, \label{subeq. the transformed problem V-RPTRMOP} \\
\text{s.t.} \ & \eqref{subeq. RPTRMOP C1}, \eqref{subeq. RPTRMOP C2}, \eqref{subeq. RPTRMOP C3},\notag
\end{align}
\end{subequations}
\noindent where $\alpha_{1}$, $\alpha_{2}$, $\alpha_{3}$, and $\alpha_{4}$ are weighting factors to balance the four Parts. Moreover, $N_{\mathrm{L}}'$ is the number of communication links in the optimized network $G'$. Due to the particularity of the post-disaster relief network, Parts III and IV are hard conditions items and their triggering should be minimized.

\par The aforementioned problem is a large-scale optimization problem due to the large number of decision variables and constraints, as well as the complexity of the objective function. First, the decision variables of the problem include the placement and excitation current weights of each UAV in swarms for relaying data. As a result, the number of decision variables increases rapidly as the network size expands, particularly when there are a large number of UAVs and links. Moreover, the number of constraints grows proportionally with the number of UAV swarms, communication links, and UAVs in each swarm. In addition, for large networks with numerous communication links, optimizing this objective function becomes computationally expensive. Therefore, given the significant number of decision variables, the constraints, and the computational complexity of the objective function, the transformed problem Eq. \eqref{eq. V-RPTRMOP} is qualified as a large-scale optimization problem.

\par Then, we characterize the NP-hardness and non-convexity of the V-RPTRMOP in the following theorems.

\begin{Theorem}
\label{Theorem: NP-hard}
The problem in Eq. \eqref{eq. V-RPTRMOP} is NP-hard since it is a nonlinear multi-dimensional 0-1 knapsack problem.
\end{Theorem}
\emph{Proof:} For simplicity, we assume the placement of each UAV is pre-designed and fixed. In this case, the excitation current weights for all UAVs $\mathbf{I}$ become the exclusive decision variable. Accordingly, the V-RPTRMOP is as follows:
\begin{subequations}
\label{eq. multi-dimensional 0-1 V-RPTRMOP}
\begin{align}
\underset{\mathbf{I}}{\text{min}} & \quad f, \label{subeq. subproblem function} \\
\text{s.t.}  \ & C1: I_{u}^{s} \in \{0,1\}, \forall s\in \mathcal{U}, u\in \mathcal{S}_{s}, \label{subeq. C1 in simplified subproblem 3}\\
& C2:\sum\nolimits_{u=1}^{N_{\mathrm{U}}}I_{u}^{s}<N_{\mathrm{U}}, \forall s\in \mathcal{U}, u\in \mathcal{S}_{s}.
\end{align}
\end{subequations}

\par It is shown that the simplified V-RPTRMOP is a nonlinear multi-dimensional 0-1 knapsack problem which has shown to be a typical NP-hard problem~\cite{GOOS2020201knapsack}. Thus, the simplified V-RPTRMOP is NP-hard, and the V-RPTRMOP is also proven to be NP-hard accordingly.
{\hfill $\blacksquare$\par}

\begin{Theorem}
\label{Theorem: non-convex}
The transformed problem in Eq. \eqref{eq. V-RPTRMOP} is non-convex since each Part of the objective function in Eq.\eqref{subeq. the transformed problem V-RPTRMOP} is non-convex.
\end{Theorem}
\emph{Proof:} The objective function of the V-RPTRMOP consists of four Parts. In Part I, the  absolute value function $|R_{i,j} - \overline{R}_{i,j}|$ is non-convex since it is not differentiable at $R_{i,j} = \overline{R}_{i,j}$. Moreover, the square root function $\sqrt{\sum_{\langle i, j \rangle \in \mathcal{E}}|R_{i,j} - \overline{R}_{i,j}|^{2}/N_{\mathrm{L}}'}$ in Part II is non-linear and concave. Finally, the penalty terms $\delta_{i,j}$ and $\eta$ in Parts III and IV take values of either 0 or 1, making them discontinuous. Therefore, as all four Parts of the objective function are non-convex, the overall objective function in Eq. \eqref{subeq. the transformed problem V-RPTRMOP} is non-convex.
Consequently, the transformed problem in Eq. \eqref{eq. V-RPTRMOP} is non-convex as well.
{\hfill $\blacksquare$\par}

\par For the V-RPTRMOP, which is large-scale, NP-hard, and non-convex, potential solution methods include convex optimization, DRL, and heuristic algorithms. However, due to the complexity of the objective function and constraints, it is impractical to convert the problem into a convex optimization problem. Therefore, convex optimization is not suitable for solving the V-RPTRMOP. Moreover, DRL models require extensive training time and incur significant overhead, rendering them unsuitable for rapidly changing post-disaster environments. Given these limitations, heuristic algorithms, commonly used for optimization problems involving beamforming techniques~\cite{Jayaprakasam2017nsga2},~\cite{li2020network}, present a feasible solution, which motivate us to propose a heuristic algorithm to solve the V-RPTRMOP.

\subsection{Diffusion Model-enabled Particle Swarm Optimization}
\label{subsec: DM-PSO}

\noindent Traditional PSO is highly effective in exploitation, refining solutions by iteratively updating the positions of particles based on their individual and global best positions. However, it often faces challenges in exploration, particularly in high-dimensional solution spaces. Diffusion models (DM) excel at exploring the solution space by generating diverse potential candidates, thereby enhancing global exploration. Thus, a DM-PSO is proposed to tackle the transformed V-RPTRMOP, which couples the exploitation efficiency of PSO and the exploration strength of diffusion models. Specifically, the main framework of the proposed algorithm is presented in Algorithm \ref{algo: Diffusion-based Perturbation Scheme}, and the key steps are described as follows:

\par a) \emph{Initialization Stage.} Initialize the velocity and position of each agent randomly, and calculate the objective function value for position of each agent. Then, update the global and local bests based on the obtained objective function values.

\par b) \emph{Update Stage.} During each iteration, the velocity $\mathbf{v}_{i}$ of the $i$th agent is updated as follows~\cite{Kennedy1995PSO}:
\begin{equation}
\mathbf{v}_{i} = \omega \mathbf{v}_{i} + c_{1}r_{1}(\mathbf{x}_{\mathrm{pbest},i}-\mathbf{x}_{i}) + c_{2}r_{2}(\mathbf{x}_{\mathrm{gbest}}-\mathbf{x}_{i}),
\label{eq. velocity update}
\end{equation}
\noindent where $\omega$ is the inertia weight, $c_{1}$ and $c_{2}$ refer to the learning coefficients, $r_{1}$ and $r_{2}$ are random numbers, subject to (0,1). Moreover, $\mathbf{x}_{\mathrm{pbest},i}$ and $\mathbf{x}_{\mathrm{gbest}}$ represent the locations of personal and global best agents, respectively.

\begin{algorithm}[htbp]
    \caption{DM-PSO}
    \label{algo: generative_model_based_PSO}
    \KwIn{Population size $N_{\mathrm{pop}}$, maximum iterations $T_{\mathrm{max}}$, objective function $f$.} 
    \KwOut{Position $\mathbf{x}_{\mathrm{gbest}}$ and the objective function value $f_{\text{gbest}}$ of the global best.}
    
    \text{/* Initialization*/} \\
    $\mathbf{x}_{\mathrm{gbest}}$  $\gets$ $\varnothing$, $f_{\mathrm{gbest}}$  $\gets$ $\varnothing$;\\
    \For{$i = 1$ to $N_{\mathrm{pop}}$}
    {
        Initialize position $\mathbf{x}_i$ and velocity $\mathbf{v}_i$ of the $i$-th agent;\\
        Calculate the objective function value $f_i$ of $\mathbf{x}_i$;\\
        $\mathbf{x}_{\mathrm{pbest},i} \gets \mathbf{x}_i$, $f_{\mathrm{pbest},i}$  $\gets$ $f_i$;\\
        \If{$f_i$ $<$ $f_{\mathrm{gbest}}$}
        {
            $\mathbf{x}_{\mathrm{gbest}}$  $\gets$ $\mathbf{x}_i$, $f_{\mathrm{gbest}}$  $\gets$ $f_i$;\\
        }
    }
    \For{$t = 1$ to $T_{\mathrm{max}}$}
    {
        \text{/* Update*/} \\
        \For{$i = 1$ to $N_{\mathrm{pop}}$}
        {
            Update $\mathbf{x}_i$ and $\mathbf{v}_i$ of the $i$-th agent based on Eqs. \eqref{eq. velocity update} and \eqref{eq. position update};\\
            Calculate the objective function value $f_i$ of $\mathbf{x}_i$;\\
            \If{$f_i$ $<$ $f_{\mathrm{pbest},i}$}
            {
                $\mathbf{x}_{\mathrm{pbest},i} \gets \mathbf{x}_i$, $f_{\mathrm{pbest},i}$  $\gets$ $f_i$;\\
                \If{$f_{\mathrm{pbest},i}$ $<$ $f_{\mathrm{gbest}}$}
                {
                    $\mathbf{x}_{\mathrm{gbest}}$  $\gets$ $\mathbf{x}_i$, $f_{\mathrm{gbest}}$  $\gets$ $f_i$;\\
                }
            }
        }
        \text{/* Diffusion*/} \\
        Update $\mathbf{x}$ based on Algorithm \ref{algo: Diffusion-based Perturbation Scheme};\\
    }
    
    Return $f_{\text{gbest}}$.
\end{algorithm}

\par According to the obtained $\mathbf{v}_{i}$, the position $\mathbf{x}_{i}$ of the $i$th agent is updated as follows:
\begin{equation}
\mathbf{x}_{i} = \mathbf{x}_{i} + \mathbf{v}_{i}.
\label{eq. position update}
\end{equation}

\par c) \emph{Diffusion Stage.} To avoid premature convergence and maintain the diversity of agents, we introduce a diffusion-enabled perturbation scheme. Specifically, the framework is outlined in Algorithm \ref{algo: Diffusion-based Perturbation Scheme} and the details are as follows:

\par \uline{Crowding distance calculation}. \emph{First}, for the $p$th agent, the scheme calculates $\mathbf{m}_p = \{d_{p, 1}, ..., d_{p, N_{\mathrm{pop}}}\}$ which contains the Euclidean distances from the $p$th agent to other agents. \emph{Second}, the scheme sorts the distance vector $\mathbf{m}_{p}$ into $\mathbf{m}_{p}^{\mathrm{sorted}}=\{d_{p, \sigma_{1}}, ..., d_{p, \sigma_{N_{\mathrm{pop}}}}\}$ in ascending order. \emph{Third}, the first $N_{\mathrm{N}}$ agents denoted as $\sigma_1, ..., \sigma_{N_{\mathrm{N}}}$ are selected as the neighbors of the $p$th agent. \emph{Finally}, the crowding distance of the $p$th agent is calculated as follows:
\begin{equation}
cd_{p} = \sum\nolimits_{j=1}^{N_{\mathrm{N}}} d_{p, \sigma_{j}} / N_{\mathrm{N}}.
\label{eq. crowding distance}
\end{equation}
\noindent Then, $cd_{p}$ is put into the crowding distance vector $\mathbf{cd}$. Finally, the scheme sorts $\mathbf{cd}$ into $\mathbf{cd}^{\mathrm{sorted}}$ in ascending order.

\par \uline{Agent position perturbation}. \emph{First}, the number of perturbed agents $N_{\mathrm{P}}$ is calculated as follows:
\begin{equation}
N_{\mathrm{P}} = N_{\mathrm{pop}}\left[p_{\mathrm{max}} - (p_{\mathrm{max}}-p_{\mathrm{min}})t/T_{\mathrm{max}}\right].
\label{eq. calculation of Nd}
\end{equation}
\noindent \emph{Second}, the first $N_{\mathrm{P}}$ agents in $\mathbf{cd}^{\mathrm{sorted}}$, corresponding to those with the smallest crowding distances, are selected for perturbation. These agents are denoted as $\mu_1, ..., \mu_{N_{\mathrm{P}}}$. \emph{Third}, perturbation is conducted by fine-tuning the forward process of the diffusion model. Specifically, for each of the selected agents $\mu_p \ (\text{for}\ p = 1, ..., N_{\mathrm{P}})$, noise is added to their current position $\mathbf{x}_{\mu_p}$ at each time step as follows:
\begin{equation}
{\mathbf{x}_{\mu_p}}_{s} =\sqrt{\alpha_{\mathrm{s}}} \mathbf{x}_{\mu_p} + \sqrt{1-\alpha_{\mathrm{s}}} \mathbf{z}_{\mathrm{s}},\quad \mathbf{z}_{\mathrm{s}} \sim \mathcal{N}(0, I),
\label{eq. add noise}
\end{equation}

\noindent where $\alpha_{\mathrm{s}}$ is a coefficient that controls the weight of the original data and noise, and $\mathbf{z}_{\mathrm{s}}$ is noise sampled from normal distribution. Based on this, the updated $\mathbf{x}$ is obtained.

\par Once upon $T_{\mathrm{max}}$ is reached, the DM-PSO is suspended, and the position $\mathbf{x}_{\mathrm{gbest}}$ along with the objective function value $f_{\text{gbest}}$ of the global best are obtained. Otherwise, the proposed algorithm continues as scheduled.

\begin{algorithm}[htbp]
    \caption{Diffusion-enabled Perturbation Scheme}
    \label{algo: Diffusion-based Perturbation Scheme}
    \KwIn{Population size $N_{\mathrm{pop}}$, positions $\mathbf{x}$, neighbors $N_{\mathrm{N}}$, maximum iterations $T_{\mathrm{max}}$, maximum $p_{\mathrm{max}}$ and minimum $p_{\mathrm{min}}$ proportion of perturbed agents.}
    \KwOut{Updated positions $\mathbf{x}$.}
    Initialize crowding distance vector $\mathbf{cd}$;\\
    \For{$p = 1$ \textbf{to} $N_{\mathrm{pop}}$}
    {
        Calculate distance vector $\mathbf{m}_p$ for the $p$th agent;\\
        Sort $\mathbf{m}_p$ into $\mathbf{m}_p^{\mathrm{sorted}}$ in ascending order;\\
        Select $N_{\mathrm{N}}$ closest neighbors as $\sigma_1, ..., \sigma_{N_{\mathrm{N}}}$;\\
        Compute crowding distance $cd_{p}$ using Eq. \eqref{eq. crowding distance};\\
        Update $\mathbf{cd} \gets \mathbf{cd} \cup cd_{p}$;
    }
    Sort $\mathbf{cd}$ into $\mathbf{cd}^{\mathrm{sorted}}$ in ascending order;\\
    Compute the number of perturbed agents $N_{\mathrm{P}}$ using Eq. \eqref{eq. calculation of Nd};\\ 
    Select the first $N_{\mathrm{P}}$ agents in $\mathbf{cd}^{\mathrm{sorted}}$ as $\mu_1, ... \mu_{N_{\mathrm{P}}}$;\\
    \For{$p = 1$ \textbf{to} $N_{\mathrm{P}}$}
    {
        Add noise to position $\mathbf{x}_{\mu_p}$ of the $\mu_p$th agent using Eq. \eqref{eq. add noise};\\
    }
    Return $\mathbf{x}$.
\end{algorithm}

\par Given the aforementioned general framework of the DM-PSO, we analyze the computational complexity of the algorithm in the following theorem.

\vspace{1mm}
\begin{Theorem}
\label{Theorem: complexity analysis}
The complexity of the proposed DM-PSO is $\mathcal{O}\left(T_{\mathrm{max}}\left(N_{\mathrm{L}}' N_{\mathrm{U}} N_{\mathrm{pop}} + N_{\mathrm{pop}}^2 \log N_{\mathrm{pop}}\right)\right)$.
\end{Theorem}
\emph{Proof:} For ease of presentation, the complexity of each part of the algorithm is first analyzed as follows:

\begin{itemize}

\item \emph{Initialization Stage}. The computational complexity of calculating the objective function value once is $\mathcal{O}\left(N_{\mathrm{L}}' N_{\mathrm{U}}\right)$. As the computational complexity of the initialization stage is primarily determined by the calculation of the objective function, the overall computational complexity of the initialization stage is $\mathcal{O}\left(N_{\mathrm{L}}' N_{\mathrm{U}} N_{\mathrm{pop}}\right)$.

\item \emph{Update Stage}. Similar to the initialization stage, the most computationally expensive part of this stage is also the calculation of the objective function, with a complexity of $\mathcal{O}\left(T_{\mathrm{max}} N_{\mathrm{L}}' N_{\mathrm{U}} N_{\mathrm{pop}}\right)$.

\item \emph{Diffusion Stage}. The computational complexity of this stage is mainly determined by the crowding distance vector calculation and sort. First, the complexity of calculating crowding distance for each agent involves computing distance vector and sorting, with complexities of $\mathcal{O}\left(N_{\mathrm{pop}}\right)$ and $\mathcal{O}\left(N_{\mathrm{pop}} \log N_{\mathrm{pop}}\right)$, respectively. Thus, the complexity of crowding distance vector calculation is $\mathcal{O}\left(N_{\mathrm{pop}}^2 \log N_{\mathrm{pop}}\right)$. Moreover, the complexity of sorting the crowding distance vector is $\mathcal{O}\left(N_{\mathrm{pop}} \log N_{\mathrm{pop}}\right)$. Therefore, the overall complexity of the diffusion stage is $\mathcal{O}\left(T_{\mathrm{max}} N_{\mathrm{pop}}^2 \log N_{\mathrm{pop}}\right)$.

\end{itemize}

\begin{figure*}[htbp]
\centering
\subfigure[]{
\includegraphics[width=0.3\linewidth]{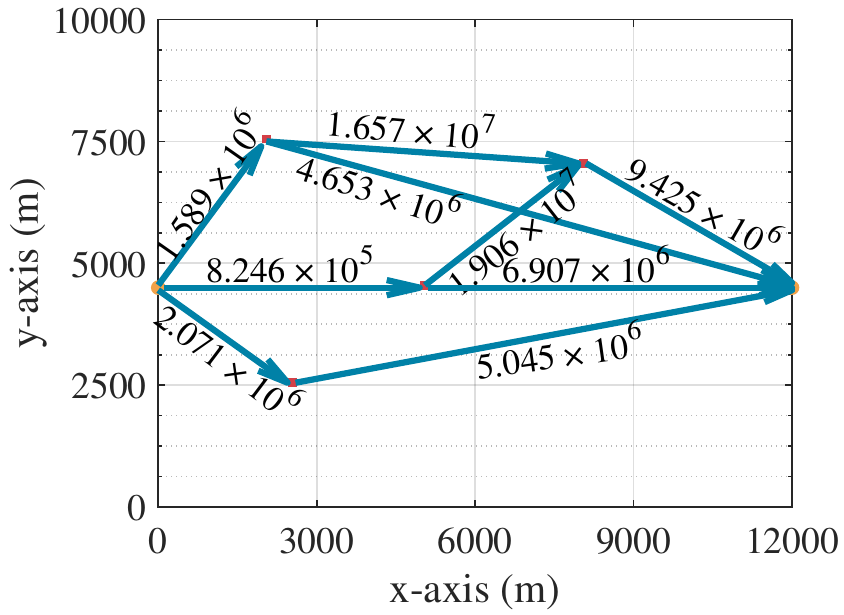}
}
\subfigure[]{
\includegraphics[width=0.3\linewidth]{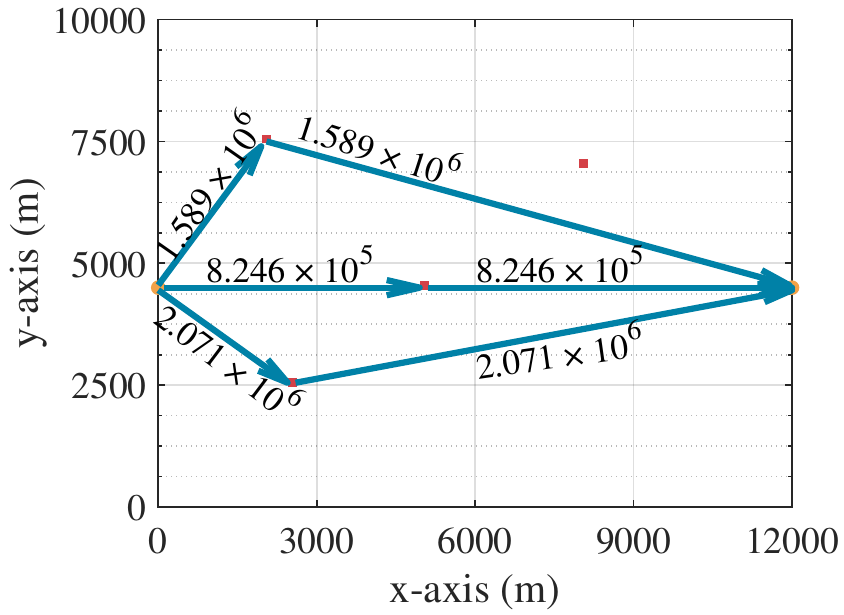}
}
\subfigure[]{
\includegraphics[width=0.3\linewidth]{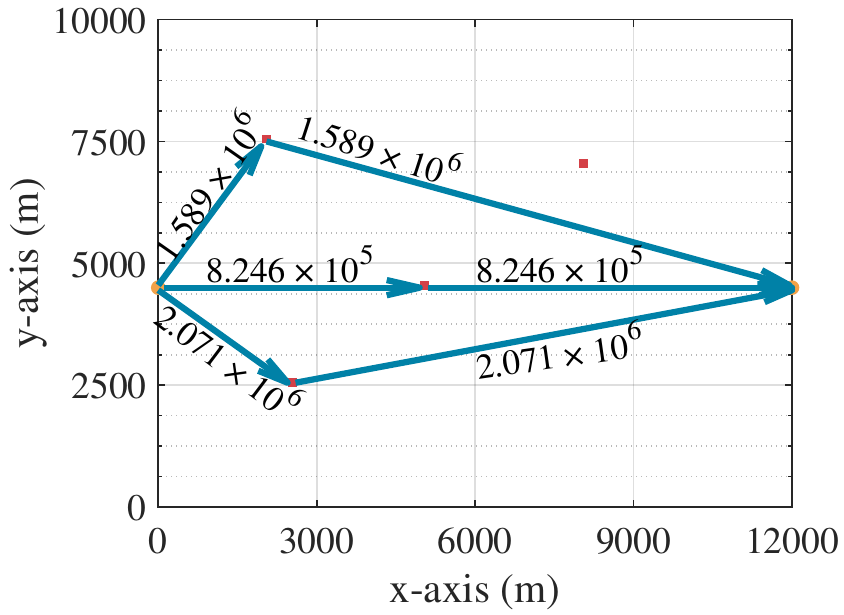}
}
\caption{The communication networks in different stages for case 1. (a) The communication network in the first stage after deducing the theoretical upper bound on the transmission rate of each link. (b) The optimized network in the first stage after conducting Ford-Fulkerson algorithm, with the theoretical upper bound on the transmission rate of the whole network is $4.48 \times 10^{6}$ bps. (c) The actual communication network in the second stage after UAV swarm optimization by DM-PSO, with the actual transmission rate of the whole network is $4.48 \times 10^{6}$ bps.}
\label{fig: optimization results of the proposed approach for case 1}
\end{figure*}

\begin{figure*}[htbp]
\centering
\subfigure[]{
\includegraphics[width=0.30\linewidth]{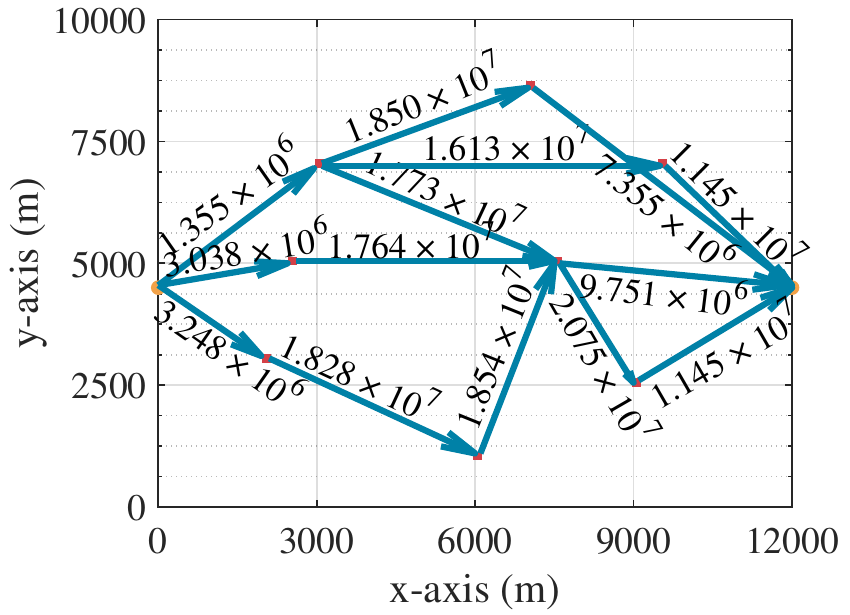}
}
\subfigure[]{
\includegraphics[width=0.30\linewidth]{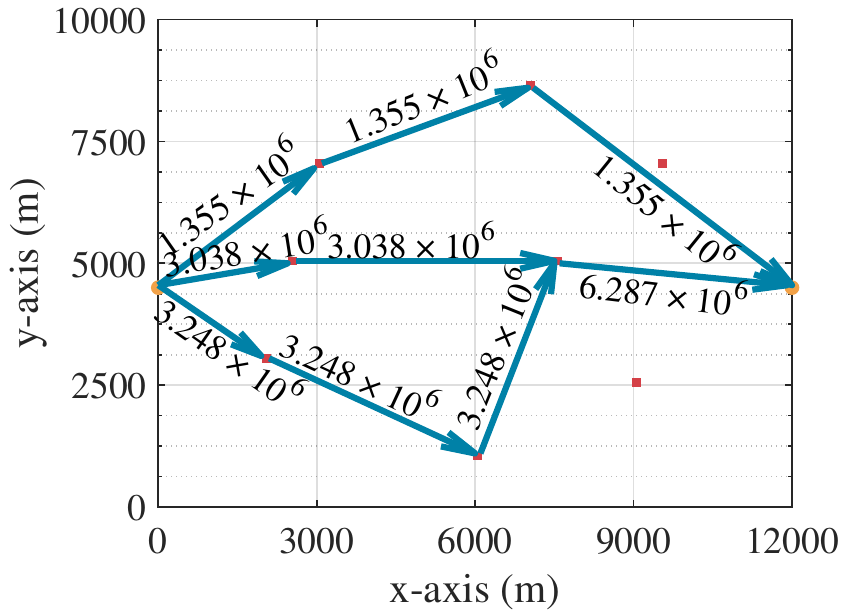}
}
\subfigure[]{
\includegraphics[width=0.30\linewidth]{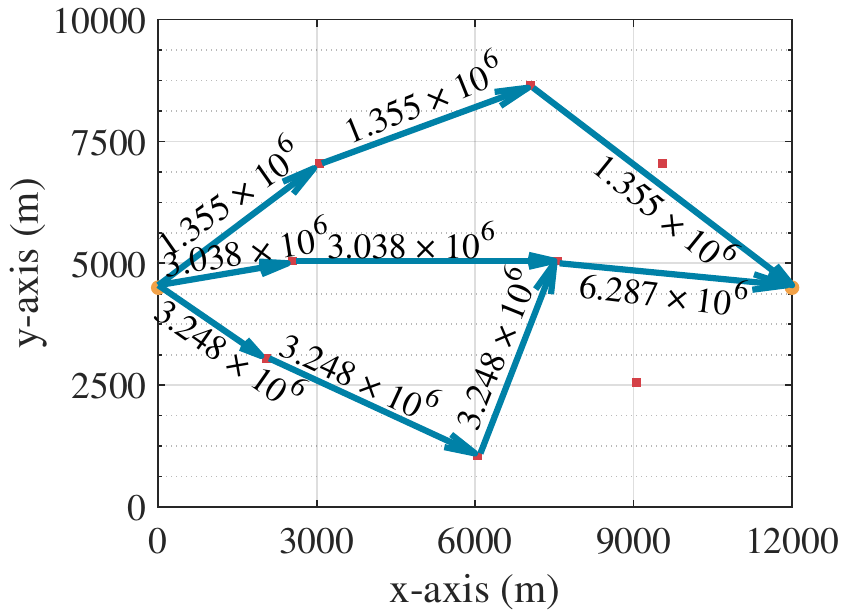}
}
\caption{The communication networks in different stages for case 2. (a) The communication network in the first stage after deducing the theoretical upper bound on the transmission rate of each link. (b) The optimized network in the first stage after conducting Ford-Fulkerson algorithm, with the theoretical upper bound on the transmission rate of the whole network is $7.64 \times 10^{6}$ bps. (c) The actual communication network in the second stage after UAV swarm optimization by DM-PSO, with the actual transmission rate of the whole network is $7.64 \times 10^{6}$ bps.}
\label{fig: optimization results of the proposed approach for case 2}
\end{figure*}

\par Accordingly, the complexity of the proposed algorithm is $\mathcal{O}\left(T_{\mathrm{max}}\left(N_{\mathrm{L}}' N_{\mathrm{U}} N_{\mathrm{pop}} + N_{\mathrm{pop}}^2 \log N_{\mathrm{pop}}\right)\right)$.
{\hfill $\blacksquare$\par}

\section{Simulation Results}
\label{sec: Simulation Results}

\noindent In this section, we execute simulations to show the advantages of deploying a UAV-swarm enabled collaborative self-organizing network to assist post-disaster communications, and validate the performance of the proposed two-stage optimization approach in solving the formulated RPTRMOP.

\subsection{Simulation Setup}
\label{subsec: Simulation Setup}

\subsubsection{Scenario Setup}
\label{subsubsec: Scenario Setup}

\noindent In this work, the locations of the ground device and the remote AP are set at (0, 4500, 0) m and (12000, 4500, 0) m, respectively. Moreover, we consider two cases where the numbers of UAV swarms are 4 and 8, respectively, with each swarm consisting of 8 UAVs. Note that the number of UAVs in each swarm can be generalized to be different. In addition, the movement area of UAVs in each swarm is defined as 100 m $\times$ 100 m, and the altitude of each UAV is constrained to [100, 120] m~\cite{Li2024TMC}. 

\subsubsection{Parameter Setting}
\label{subsubsec: Parameters Setting}

\noindent The wavelength $\lambda$, frequency $f_\mathrm{c}$, path loss exponent $\alpha$, bandwidth $B$, noise power spectral density, and transmit power of the ground device or a single UAV $p_t$ are 0.33 m, 0.9 GHz, 2~\cite{Zeng2019energyminimization}, 2 MHz~\cite{Li2024TMC}, -157 dBm/Hz~\cite{Li2024TMC}, and 0.1 W~\cite{Gupta2023power}, respectively. The threshold of the distance between any two UAVs in the same swarm $D_{\mathrm{min}}$ is 0.5 m. Moreover, the parameters for the probability LoS model $m$, $n$, $\eta^{\mathrm{LoS}}$, and $\eta^{\mathrm{NLoS}}$ are set as 4.88, 0.43~\cite{Samir2022LoS}, 0.977 and 0.00974, respectively.

\par The weighting factors of the transformed V-RPTRMOP $\alpha_1$, $\alpha_{2}$, $\alpha_{3}$, and $\alpha_{4}$ are 1, 1, $2 \times 10^7$, and $2 \times 10^7$, respectively. Note that the values of $\alpha_{3}$ and $\alpha_{4}$ are set to be relatively large since the corresponding Parts III and IV represent hard constraints. Moreover, the population size $N_{\mathrm{pop}}$, the maximum number of iterations $T_{\mathrm{max}}$, inertia weight $\omega$, and learning coefficients $c_1$ and $c_2$ are set as 100, 500, 1, 1.5, and 2, respectively. In addition, the number of agent neighbors $N_{\mathrm{N}}$, the maximum and minimum proportion of the perturbed agents $p_{\mathrm{max}}$ and $p_{\mathrm{min}}$ are 10, 0.3 and 0.1, respectively.

\subsection{Performance Evaluation}
\label{subsec: Performance Evaluation}

\subsubsection{Visualization Results of the Proposed Two-stage Optimization Approach}
\label{subsubsec: Visualization Results}

\noindent Figs. \ref{fig: optimization results of the proposed approach for case 1} and \ref{fig: optimization results of the proposed approach for case 2} illustrate the communication networks in different stages for the considered two cases. First, compared with the constructed network shown in Figs. \ref{fig: optimization results of the proposed approach for case 1}(a) and \ref{fig: optimization results of the proposed approach for case 2}(a), several links and UAV swarms are excluded from the data transmission process in Figs. \ref{fig: optimization results of the proposed approach for case 1}(b) and \ref{fig: optimization results of the proposed approach for case 2}(b). On the one hand, eliminating redundant communication links reduces the waste of communication resources. On the other hand, UAV swarms that do not participate in the whole data transmission process can execute other relief missions in the post-disaster area, significantly enhancing the efficiency of disaster relief. Second, the optimized transmission rates of several links in Figs. \ref{fig: optimization results of the proposed approach for case 1}(b) and \ref{fig: optimization results of the proposed approach for case 2}(b) are lower than their theoretical upper bounds, indicating that UAVs are not always required to operate at maximum transmission power when transmitting data, which also helps conserve their limited resources. Finally, it can be observed from Figs. \ref{fig: optimization results of the proposed approach for case 1}(c) and \ref{fig: optimization results of the proposed approach for case 2}(c) that the transmission rate of each communication link is well optimized, ensuring no congestion occurs. Moreover, the difference between the actual transmission rates of the communication links in Fig. \ref{fig: optimization results of the proposed approach for case 1}(c) and the expected rates in Fig. \ref{fig: optimization results of the proposed approach for case 1}(b) is minimal, and the actual transmission rate of the whole network closely aligns with the theoretical upper bound, which can be attributed to the first three proposals of V-RPTRMOP in Section \ref{sec: Step Two}. Note that this analysis also applies to Fig. \ref{fig: optimization results of the proposed approach for case 2}.

\par In conclusion, the proposed two-stage optimization approach can efficiently optimize the transmission rate of the entire network and is resource-friendly for post-disaster communications.

\subsubsection{Comparison of Traffic Routing Design with Other Existing Protocols}
\label{subsubsec: The Proposed Routing Design vs. Existing Routing Protocols}

\noindent To evaluate the proposed traffic routing design, several existing routing protocols are used as benchmarks, detailed as follows:

\begin{itemize}

\item \emph{Routing Information Protocol (RIP)}. A distance-vector routing protocol that selects the path with the fewest hops as the optimal routing~\cite{hedrick1988routing}.

\item \emph{Open Shortest Path First (OSPF)}. A link-state protocol which uses the Dijkstra algorithm to determine the traffic routing from the ground device to the AP~\cite{pioro2002open}.

\item \emph{Greedy Perimeter Stateless Routing (GPSR)}. A location-based protocol where each node selects the neighbor closest to the location of AP as the next hop~\cite{karp2000gpsr}.
    
\item \emph{Zone Routing Protocol (ZRP)}. A hybrid protocol that defines areas by hop count, using proactive routing within zones and on-demand routing between them~\cite{beijar2002zone}.

\item \emph{Temporally Ordered Routing Algorithm (TORA)}. An on-demand routing protocol for dynamic networks, where each node selects the neighbor with the lowest height value as the next hop~\cite{park1998performance}.

\end{itemize}

\begin{figure}[htbp]
\centering
\subfigure[Case 1]{\includegraphics[width=0.45\linewidth]{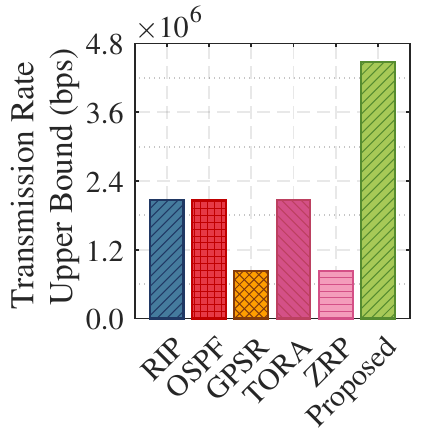}}
\subfigure[Case 2]{\includegraphics[width=0.45\linewidth]{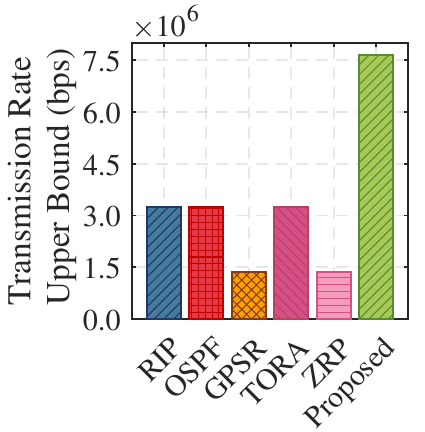}}
\caption{Theoretical upper bound on the transmission rate of the whole network obtained by several existing routing protocols and the proposed traffic routing method for the two cases.}
\label{fig. routing design}
\end{figure}

\begin{figure}[htbp]
\centering
\subfigure[Case 1]{\includegraphics[width=0.78\linewidth]{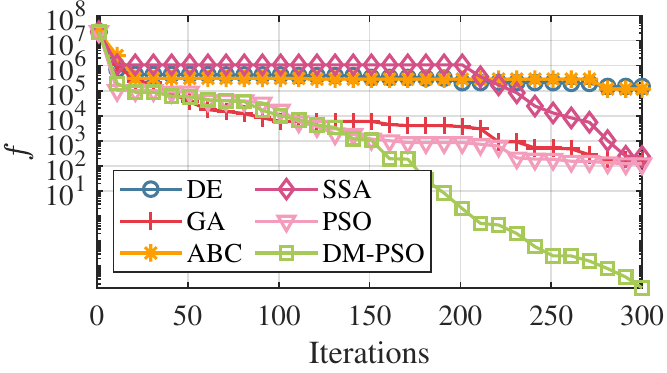}}
\quad
\subfigure[Case 2]{\includegraphics[width=0.78\linewidth]{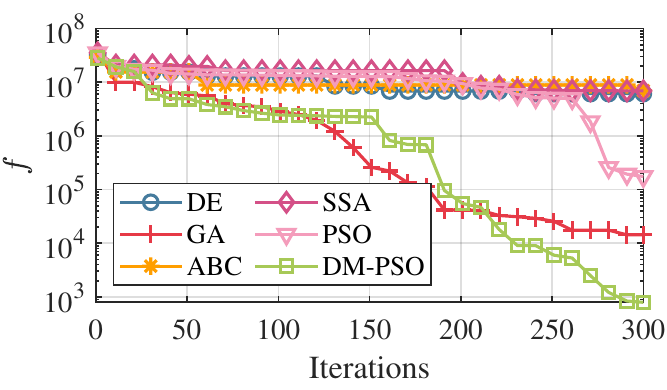}}
\caption{Objective function value $f$ of the V-RPTRMOP obtained by several benchmark heuristic algorithms and the proposed DM-PSO for the two cases.}
\label{fig. convergence}
\end{figure}

\par As shown in Fig. \ref{fig. routing design}, there is a clear performance hierarchy, with GPSR and ZRP performing the worst among the adopted protocols. Moreover, RIP, OSPF, and TORA demonstrate better performance than GPSR and ZRP, however, their theoretical upper bounds on transmission rates remain unsatisfactory. Finally, the proposed traffic routing method achieves the highest theoretical upper bound on the transmission rate of the whole network, which is almost 2 times that of RIP, OSPF, and TORA, and 4 times that of GPSR and ZRP, demonstrating its superior efficiency. This significant improvement stems from the fact that conventional protocols restrict each node to transmitting data to a single receiver, whereas the proposed method enables nodes to transmit data to multiple receivers simultaneously. Consequently, data can reach the AP through multiple paths, substantially enhancing the overall transmission rate.

\subsubsection{Comparison of DM-PSO with Other Algorithms}
\label{subsubsec: Comparison of DM-PSO with Other Algorithms}

\noindent We compare the performance of our proposed DM-PSO with several benchmark heuristic algorithms, including differential evolution (DE)~\cite{das2010differential}, genetic algorithm (GA)~\cite{mirjalili2019genetic}, artificial bee colony (ABC)~\cite{karaboga2009comparative}, salp swarm algorithm (SSA)~\cite{Mirjalili2017salp}, and conventional PSO. 

\par The objective function values $f$ for the V-RPTRMOP obtained by these algorithms for the considered two cases are shown in Fig. \ref{fig. convergence}. As can be seen, the curves of DE and ABC remain stable with minimal reductions in the objective function value as the number of iterations increases. The reason is that DE and ABC converge prematurely to local optimal solutions. Moreover, the performance of SSA and PSO in different cases is relatively unstable, since heuristic algorithms exhibit varying effectiveness when solving diverse problems or even the same problem under different setting. In addition, it can be found that GA outperforms the aforementioned three algorithms. However, final objective value of GA remains higher than that of DM-PSO. Finally, even in the later iterations, the objective function value achieved by DM-PSO continues to decrease, highlighting its superior exploration and exploitation capabilities. 

\begin{table}[htbp]
\centering
\caption{Numerical Optimization Results of the Four Parts Obtained by Several Benchmark Heuristic Algorithms and the Proposed DM-PSO for the Two Cases}
\label{table:numerical results of DM-PSO}
\begin{tabular}{|c|c|c|c|c|c|}
\hline
\small
\textbf{Case} & \textbf{Algorithm} & \textbf{Part I}    &  \textbf{Part II}     &  \textbf{Part III}     &  \textbf{Part IV}  \\ \hline
\multirow{6}{*}{\textbf{1}} & \textbf{DE}  & 101440.5 & 45196.2   & 0 & 0 \\  
& \textbf{GA} & 119.7 & 56.5 & 0 & 0 \\
& \textbf{ABC} & 83021.0  & 33286.3 & 0 & 0 \\
& \textbf{SSA} & 164.0 & 73.5 & 0 & 0 \\
& \textbf{PSO} & 80.9 & 40.8 & 0 & 0 \\
& \textbf{DM-PSO}  & \textbf{0} & \textbf{0} & 0 & 0 \\ \hline
\multirow{6}{*}{\textbf{2}} & \textbf{DE}  & 4983108.8 & 1151133.9 &  0 & 0 \\  
& \textbf{GA} & 11855.7 & 2473.6 & 0 & 0 \\
& \textbf{ABC} & 5832599.8 & 1151161.9 & 0 & 0 \\
& \textbf{SSA} & 5334978.4 & 1568413.9 & 0 & 0  \\
& \textbf{PSO} & 136976.9 & 38215.1 & 0 & 0 \\
& \textbf{DM-PSO} & \textbf{659.5} & \textbf{132.2} & 0 & 0 \\ \hline
\end{tabular}
\end{table}

\begin{figure}[htbp]
\centering
\subfigure[Case 1]{\includegraphics[width=0.45\linewidth]{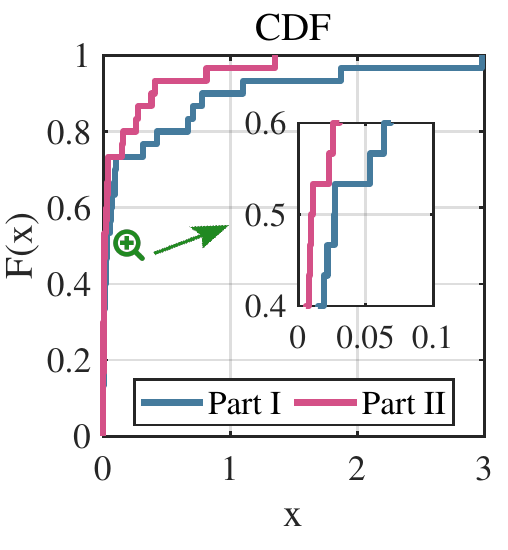}}
\subfigure[Case 2]{\includegraphics[width=0.45\linewidth]{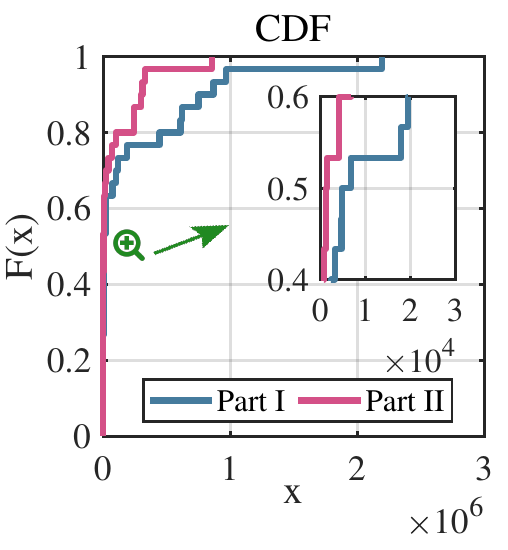}}
\caption{The CDFs of the values of Parts I and II corresponding to the solutions of DM-PSO for the two cases.}
\label{fig. stability analysis cdf}
\end{figure}

\par Table \ref{table:numerical results of DM-PSO} shows the final optimized values of the four Parts in the V-RPTRMOP obtained by several benchmark algorithms and the proposed DM-PSO for both cases 1 and 2. Intuitively, DM-PSO achieves significantly lower values in the objective function defined in Eq.~\eqref{subeq. the transformed problem V-RPTRMOP} for both Parts I and II compared with other benchmark algorithms, which further illustrates the superiority of the proposed DM-PSO. Moreover, the optimized values of Part III and Part IV are 0, the reason is that these two Parts are hard constraints and the corresponding weight factors are set relatively large, so that all algorithms adapt during the iterative process to optimize the first two Parts without triggering the latter two penalty conditions. Finally, it can be observed that, compared to the results of case 1, the values of Parts I and II obtained by these algorithms for case 2 are larger. The reason is that the network scale of case 2 is larger, with more transmission links and higher solution dimensionality during optimization, making it more challenging for these algorithms to find the optimal solution.

\par In conclusion, the performance of DM-PSO is superior to that of the other benchmark algorithms in solving the V-RPTRMOP. This improvement is due to the introduction of the diffusion model, which enables the more crowded agents to escape local optima and explore a broader search space.

\subsubsection{Stability Analysis of DM-PSO}
\label{subsubsec: Stability Analysis of the Proposed DM-PSO}

\begin{figure}[htbp]
\centering
\subfigure[Case 1]{
\includegraphics[width=0.81\linewidth]{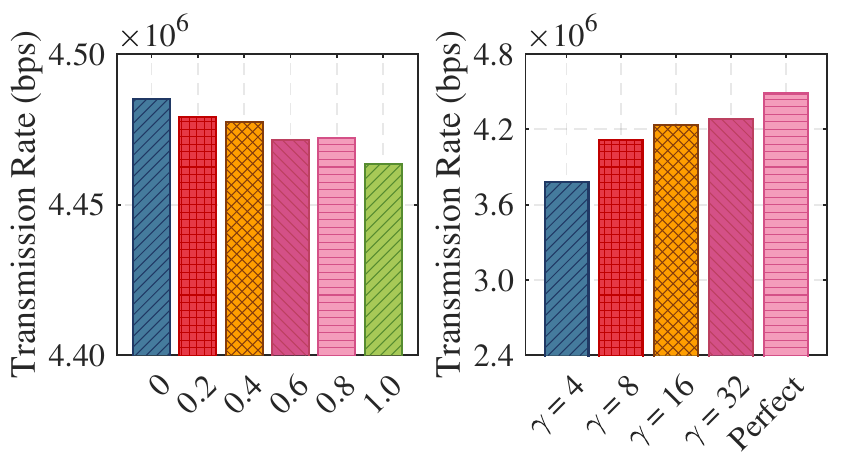}\label{fig: position jitters of 4 swarms}
}
\quad
\subfigure[Case 2]{
\includegraphics[width=0.81\linewidth]{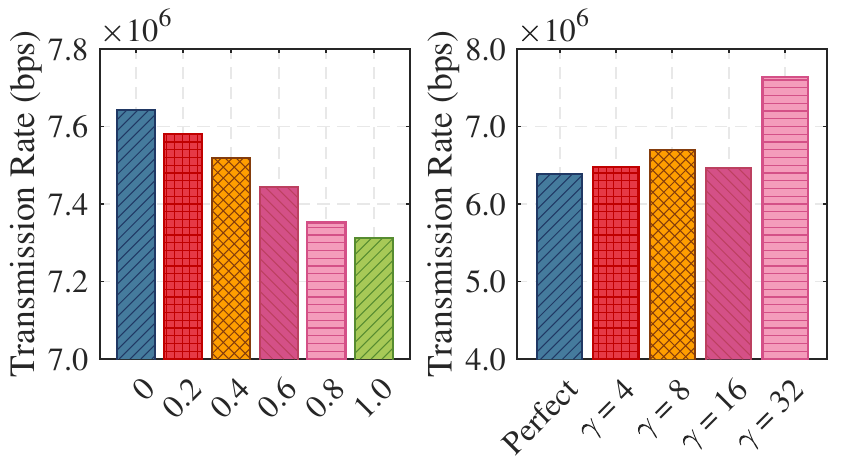}\label{fig: phase error of 4 swarms}
}
\caption{Performance analysis of the constructed communication network under two unexpected situations for the two cases.}
\label{fig: robust analysis}
\end{figure}

\noindent In this part, 30 times independently replicated trials are executed to validate the stability of the proposed DM-PSO. The cumulative distribution functions (CDFs) of the values of Parts I and II corresponding to the DM-PSO solutions are presented in Fig. \ref{fig. stability analysis cdf}, respectively. Note that Parts III and IV are not analyzed in this part, since theirs values achieved by all solutions are 0 for the two cases. As shown in Fig. \ref{fig. stability analysis cdf}(a), for case 1, 70\% of the solutions have values of Parts I and II equal to zero, while the counterparts of the remaining 30\% exhibit fluctuations within 3 units above zero. Moreover, for case 2, the values of Parts I and II corresponding to 70\% of the DM-PSO solutions remain low, while the values for the remaining  30\% are comparatively high due to the increased network scale, which makes finding the optimal solution more challenging. However, this phenomenon does not affect the transmission rate of the entire network due to the introduction of Part III. In conclusion, the stability of the proposed DM-PSO is effectively demonstrated.

\subsection{Robustness Analysis}
\label{subsec: Robustness Analysis}

\noindent The robustness of the constructed communication network is validated by evaluating the impact of two unexpected situations on network performance, which are position jitters and phase errors of UAVs, respectively. To mitigate randomness, 30 independent trials are conducted, and the average results are presented.

\begin{itemize}

\item \emph{Position Jitters of UAVs}. UAVs may deviate from their predetermined optimal locations when exposed to inclement weather conditions, resulting in inevitable deterioration of the beam pattern for each UAV swarm. To evaluate the impact of position jitters on network performance, simulations are conducted with maximum jitter values set to be 0.2 m, 0.4 m, 0.6 m, 0.8 m and 1 m, respectively~\cite{Li2015Positionerrornormaldistribution}.
    
\item \emph{Phase Errors of UAVs}. Achieving perfect phase synchronization among the collaborative UAV elements has proven to be an intractable challenge. Thus, we aim to assess the effect of imperfect phase synchronization on the actual transmission rate of the constructed network. Specifically, the AF $F_{s}^{\zeta}$ of the VAA formed by UAVs in the $s$th swarm with phase error is reformulated as~\cite{Minturn2013phase1}:
\begin{equation}
\label{eq. AF with phase error}
F_s^\zeta(\theta,\phi) = F_s(\theta,\phi)\sum\nolimits_{u=1}^{N_{\mathrm{U}}}I_u\text{e}^{j\zeta_u},
\end{equation}
\noindent where $\zeta_u$ is the phase error of the $u$th UAV, which follows the Tikhonov distribution. The probability density function of the Tikhonov distribution is referred to ~\cite{Jung2021PDF}.
\end{itemize}

\par Fig. \ref{fig: robust analysis} intuitively presents the actual transmission rate of the constructed communication network under the aforementioned unexpected situations for both cases 1 and 2. It can be observed that the actual transmission rate of the whole network will slightly decrease when position jitters occur among the UAVs for both cases. Moreover, the transmission rate of the network will deteriorate when there are phase errors among the collaborative UAVs. However, the actual transmission rate of the constructed network for case 1 is still decent even when the unexpected situations occur. Thus, the robustness of the UAV-swarm enabled collaborative self-organizing network is illustrated.

%
%
\section{Conclusion}
\label{sec: Conclusion}

\noindent In this work, we have designed a UAV-swarm enabled collaborative self-organizing network to assist post-disaster communications. We have formulated an optimization problem termed as RPTRMOP to maximize the transmission rate of the constructed communication network. Given that the formulated RPTRMOP cannot be tackled by traditional methods directly, we have proposed a two-stage optimization approach. In the first stage, the optimal traffic routing and the theoretical upper bound on the transmission rate of the network have been derived by using theoretical analysis and the Ford-Fulkerson algorithm. In the second stage, we have transformed the formulated RPTRMOP into a variant named V-RPTRMOP based on the obtained optimal traffic routing. Then, we have optimized the excitation current weight and the placement of each participating UAV via the proposed DM-PSO, so that the actual transmission rate closely can approach its theoretical upper bound. Simulation results have validated the effectiveness of the proposed two-stage optimization approach in solving the formulated problem, which demonstrates significant potential for post-disaster communications. Moreover, we have evaluated the impact of UAV position jitters and phase errors on system performance, and the decent results manifests the robustness of the constructed network.
 
\bibliographystyle{IEEEtran}
\bibliography{myref}

\vspace{-3 mm}
\begin{IEEEbiography}[{\includegraphics[width=1in,height=1.25in,clip,keepaspectratio]{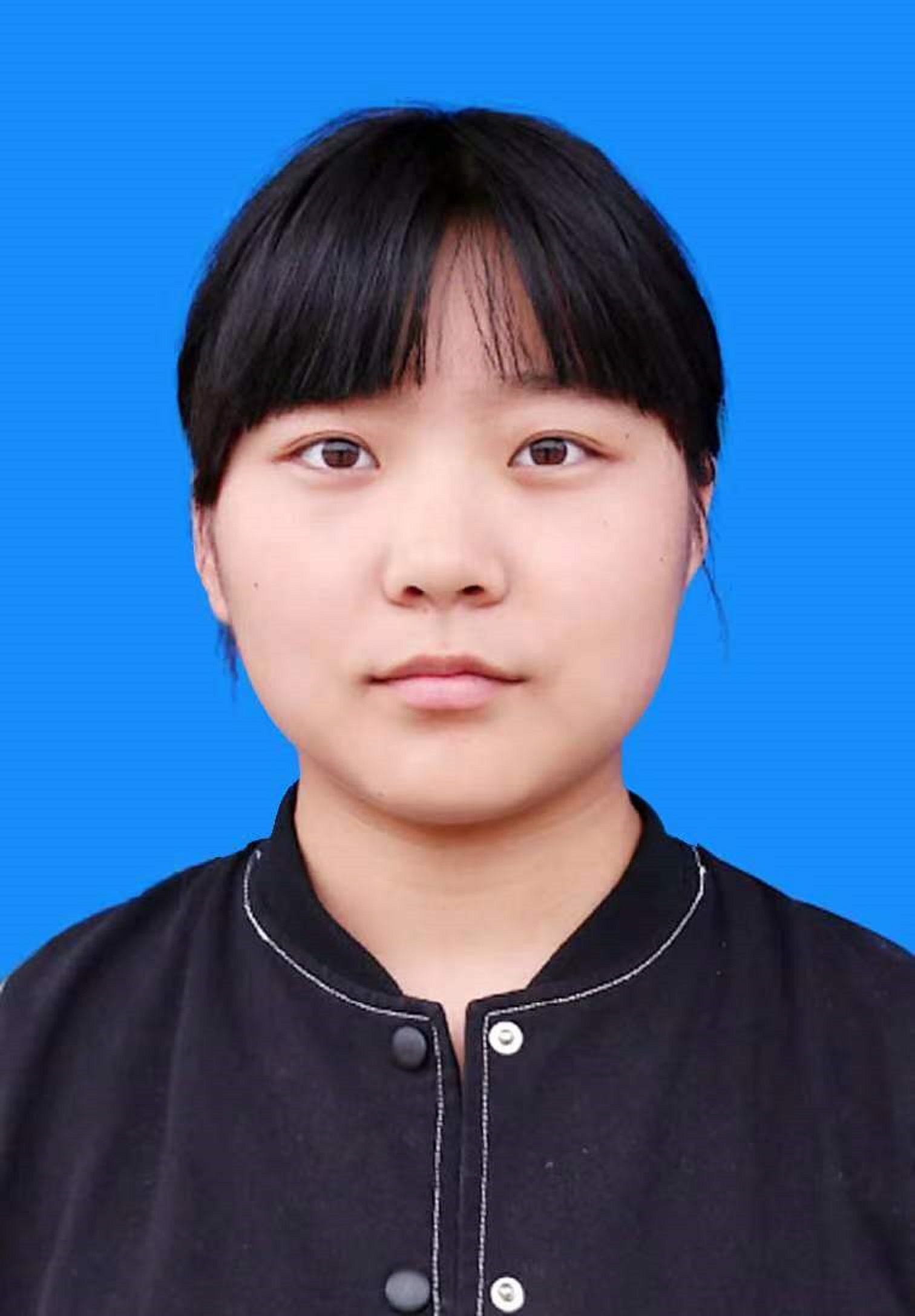}}]{Xiaoya Zheng} received the B.S. degree in software engineering from Hebei Geology University in 2021, and the M.S. degree in College of Computer Science and Technology from Jilin University in 2024. She is currently pursuing a Ph.D. degree at the College of Computer Science and Technology, Jilin University. Her research interests focus on UAV networks and optimization techniques.
\end{IEEEbiography}

\vspace{-3 mm}
\begin{IEEEbiography}[{\includegraphics[width=1in,height=1.25in,clip,keepaspectratio]{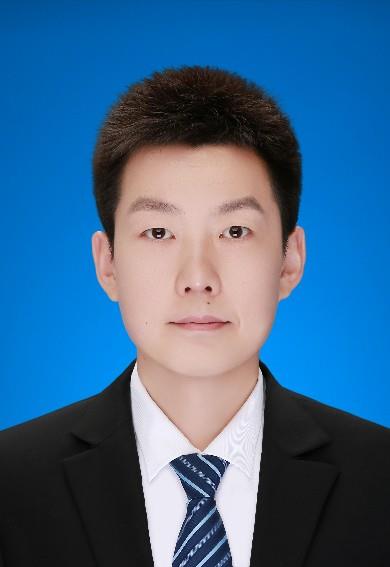}}]{Geng Sun} (Senior Member, IEEE) received the B.S. degree in communication engineering from Dalian Polytechnic University, in 2007, and the Ph.D. degree in computer science and technology from Jilin University, in 2018. He was a Visiting Researcher with the School of Electrical and Computer Engineering, Georgia Institute of Technology, USA. He is an Associate Professor in College of Computer Science and Technology at Jilin University, and his research interests include wireless networks, UAV communications, collaborative beamforming and optimizations. 
\end{IEEEbiography}

\vspace{-3 mm}
\begin{IEEEbiography}[{\includegraphics[width=1in,height=1.25in,clip,keepaspectratio]{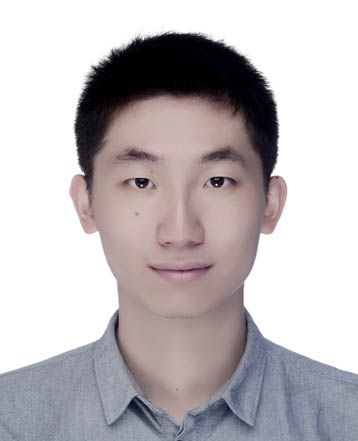}}]{Jiahui Li} (Student Member, IEEE) received the B.S. degree in Software Engineering, and the M.S. degree in Computer Science and Technology from Jilin University, Changchun, China, in 2018 and 2021, respectively. He is currently studying Computer Science at Jilin University to get the Ph.D. degree. His current research focuses on UAV networks, antenna arrays, and optimization.
\end{IEEEbiography}

\vspace{-3 mm}
\begin{IEEEbiography}[{\includegraphics[width=1in,height=1.25in,clip,keepaspectratio]{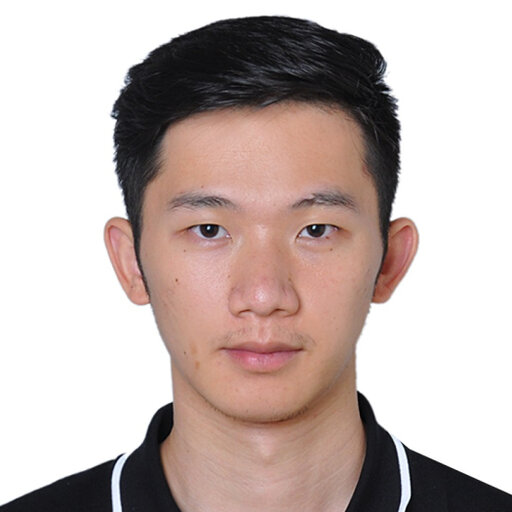}}]{Jiacheng Wang} is the research fellow in the College of Computing and Data Science at Nanyang Technological University, Singapore. Prior to that, he received the Ph.D. degree in School of Communications and Information Engineering, Chongqing University of Posts and Telecommunications, Chongqing, China. His research interests include wireless sensing, semantic communications, and generative AI, Metaverse.
\end{IEEEbiography}

\vspace{-3 mm}
\begin{IEEEbiography}[{\includegraphics[width=1in,height=1.25in,clip,keepaspectratio]{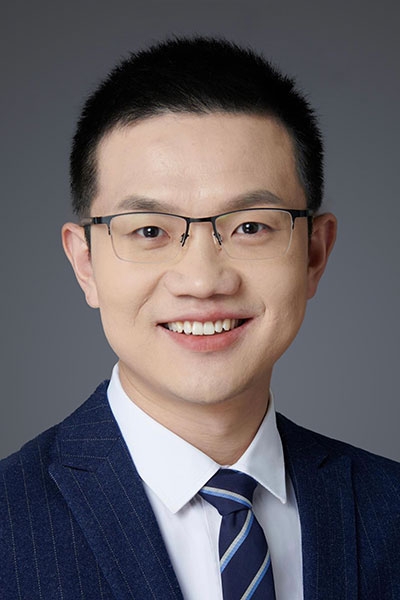}}]{Qingqing Wu}
(Senior Member, IEEE) received the B.Eng. and the Ph.D. degrees in Electronic Engineering from South China University of Technology and Shanghai Jiao Tong University in 2012 and 2016, respectively. From 2016 to 2020, he was a Research Fellow in the Department of Electrical and Computer Engineering at National University of Singapore. He is currently an Associate Professor with Shanghai Jiao Tong University. His current research interest includes IRS, UAV communications, and MIMO transceiver design. He has coauthored more than 100 IEEE journal papers with 26 ESI highly cited papers and 8 ESI hot papers, which have received more than 18,000 Google citations. He was listed as the Clarivate ESI Highly Cited Researcher in 2022 and 2021, the Most Influential Scholar Award in AI-2000 by Aminer in 2021 and World’s Top 2$\%$ Scientist by Stanford University in 2020 and 2021. 
\end{IEEEbiography}

\vspace{-3 mm}
\begin{IEEEbiography}[{\includegraphics[width=1in,height=1.25in,clip,keepaspectratio]{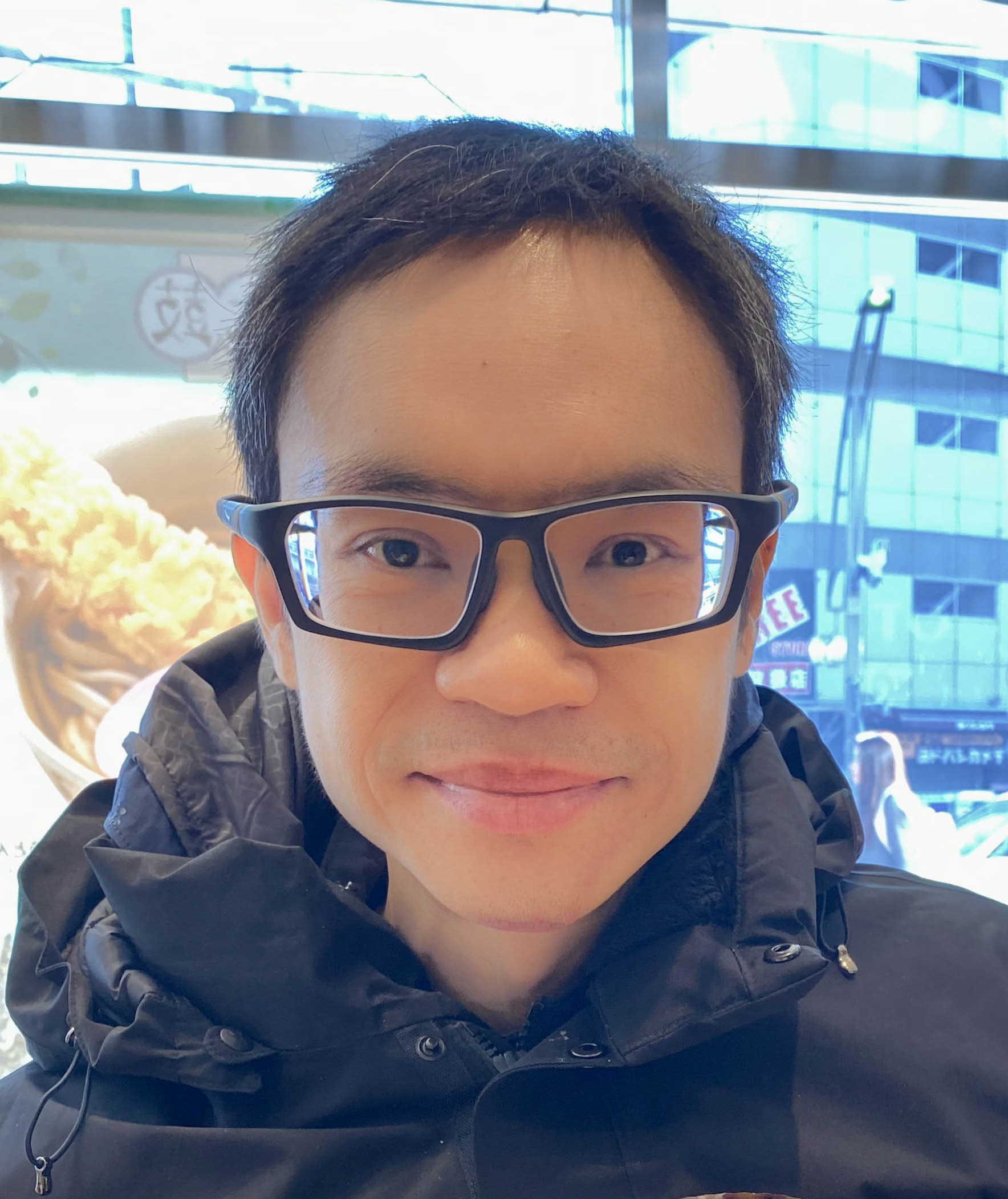}}]{Dusit Niyato}
(Fellow, IEEE) received the B.Eng.degree from the King Mongkuts Institute of Technology Ladkrabang (KMITL), Thailand, in 1999, and the Ph.D. degree in electrical and computer engineering from the University of Manitoba, Canada, in 2008. He is currently a Professor with the School of Computer Science and Engineering, Nanyang Technological University, Singapore. His research interests include the Internet of Things (IoT), machine learning, and incentive mechanism design.
\end{IEEEbiography}

\vspace{-3 mm}
\begin{IEEEbiography}[{\includegraphics[width=1in,height=1.25in,clip,keepaspectratio]{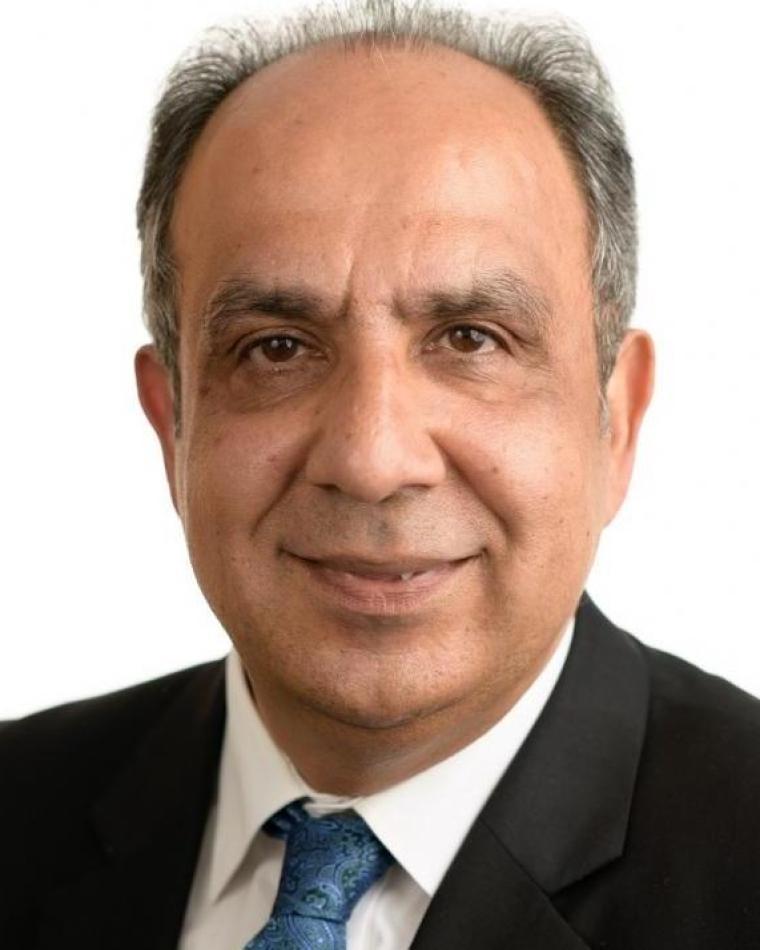}}]{Abbas Jamalipour}
(Fellow, IEEE) received the Ph.D. degree in electrical engineering from Nagoya University, Nagoya, Japan, in 1996. He is currently a Professor of ubiquitous mobile networking with The University of Sydney. He has authored nine technical books, eleven book chapters, more than 550 technical papers, and five patents, all in wireless communications and networking. He was a member of the Board of Governors of the IEEE Communications Society. He has been an Elected Member of the Board of Governors of the IEEE Vehicular Technology Society since 2014. He is a member of the Advisory Board of IEEE Internet of Things Journal. He is a fellow of the Institute of Electrical, Information, and Communication Engineers (IEICE), and the Institution of Engineers Australia, an ACM Professional Member, and an IEEE Distinguished Speaker. He was a recipient of a number of prestigious awards, such as the 2019 IEEE ComSoc Distinguished Technical Achievement Award in Green Communications, the 2016 IEEE ComSoc Distinguished Technical Achievement Award in Communications Switching and Routing, the 2010 IEEE ComSoc Harold Sobol Award, the 2006 IEEE ComSoc Best Tutorial Paper Award, and more than 15 best paper awards. He has been the General Chair or the Technical Program Chair of several prestigious conferences, including IEEE ICC, GLOBECOM, WCNC, and PIMRC. He is the Editor-in-Chief of IEEE Transactions on Vehicular Technology. He was the President of the IEEE Vehicular Technology Society from 2020 to 2021. Previously, he held the positions of the Executive Vice-President and the Editor-in-Chief of VTS Mobile World. He was the Editor-in-Chief of IEEE Wireless Communications and the Vice President-Conferences. He sits on the Editorial Board of IEEE Access and several other journals.
\end{IEEEbiography}

\end{document}